\documentclass[fleqn,usenatbib]{mnras}

\usepackage{newtxtext,newtxmath}
\usepackage[T1]{fontenc}
\usepackage{comment}
\DeclareRobustCommand{\VAN}[3]{#2}
\let\VANthebibliography\thebibliography
\def\thebibliography{\DeclareRobustCommand{\VAN}[3]{##3}\VANthebibliography}

\usepackage{graphicx}	
\usepackage{amsmath}	
\usepackage{orcidlink}

\newcommand{\NtcNightOne}{\mbox{$2460133.8485^{+0.0061}_{-0.0090}$}}

\newcommand{\NtcNightTwo}{\mbox{$2460152.7360^{+0.0013}_{-0.0014}$}}

\newcommand{\NcConNightOne}{\mbox{$1.1738^{+0.0043}_{-0.0044}$}}
\newcommand{\NcLinNightOne}{\mbox{$-0.524\pm0.012$}}
\newcommand{\NcQuadNightOne}{\mbox{$0.3819^{+0.0079}_{-0.0080}$}}
\newcommand{\NcConNightTwo}{\mbox{$1.1968\pm0.0031$}}
\newcommand{\NcLinNightTwo}{\mbox{$-0.5876^{+0.0086}_{-0.0087}$}}
\newcommand{\NcQuadNightTwo}{\mbox{$0.4278^{+0.0060}_{-0.0058}$}}

\newcommand{\tc}{$T_{\rm C}$}
\newcommand{\rprs}{\mbox{$R_{\rm P}/R_{\ast}$}}
\newcommand{\ars}{\mbox{$a/R_{\ast}$}}

\title[A possible misaligned orbit for AU\,Mic\,c]{A possible misaligned orbit for the young planet AU\,Mic\,c\thanks{Based on observations made with the ESO's Very Large Telescope at Paranal Observatory under programme 111.24VF.002. This study uses CHEOPS data observed as part of the Guaranteed Time Observervation (GTO) programme CH\_PR00071.}}

\author[H. Yu et al.]{
\parbox{\textwidth}{
H.~Yu\textsuperscript{\hyperlink{inst:1}{1}}\thanks{E-mail: \href{mailto:haochuan.yu@physics.ox.ac.uk}{haochuan.yu@physics.ox.ac.uk}.},
Z.~Garai\textsuperscript{\hyperlink{inst:2}{2},\hyperlink{inst:3}{3},\hyperlink{inst:4}{4}}\orcidlink{0000-0001-9483-2016}, 
M.~Cretignier\textsuperscript{\hyperlink{inst:1}{1}}\orcidlink{0000-0002-2207-0750}, 
Gy.~M.~Szabó\textsuperscript{\hyperlink{inst:2}{2},\hyperlink{inst:3}{3}}\orcidlink{0000-0002-0606-7930}, 
S.~Aigrain\textsuperscript{\hyperlink{inst:1}{1}}\orcidlink{0000-0003-1453-0574}, 
D.~Gandolfi\textsuperscript{\hyperlink{inst:5}{5}}\orcidlink{0000-0001-8627-9628}, 
E.~M.~Bryant\textsuperscript{\hyperlink{inst:6}{6}}\orcidlink{0000-0001-7904-4441}, 
A.~C.~M.~Correia\textsuperscript{\hyperlink{inst:7}{7}}\orcidlink{0000-0002-8946-8579},
B.~Klein\textsuperscript{\hyperlink{inst:1}{1}}\orcidlink{0000-0003-0637-5236},
A.~Brandeker\textsuperscript{\hyperlink{inst:8}{8}}\orcidlink{0000-0002-7201-7536}
% ...
}
~\\
~\\
Affiliations are listed at the end of the paper in Appendix~\ref{sec:affiliations}.
}

\date{Accepted XXX. Received YYY; in original form ZZZ}

\pubyear{2024}

\begin{document}
\label{firstpage}
\pagerange{\pageref{firstpage}--\pageref{lastpage}}
\maketitle

\begin{abstract}
The AU Microscopii planetary system is only 24 Myr old, and its geometry may provide clues about the early dynamical history of planetary systems. Here, we present the first measurement of the Rossiter-McLaughlin effect for the warm sub-Neptune AU Mic c, using two transits observed simultaneously with the European Southern Observatory's (ESO's) Very Large Telescope (VLT)/Echelle SPectrograph for Rocky Exoplanets and Stable Spectroscopic Observations (ESPRESSO), CHaracterising ExOPlanet Satellite (CHEOPS), and Next-Generation Transit Survey (NGTS). After correcting for flares and for the magnetic activity of the host star, and accounting for transit-timing variations, we find the sky-projected spin-orbit angle of planet c to be in the range $\lambda_c=67.8_{-49.0}^{+31.7}$\,degrees (1-$\sigma$). We examine the possibility that planet c is misaligned with respect to the orbit of the inner planet b ($\lambda_b=-2.96_{-10.30}^{+10.44}$\,degrees), and the equatorial plane of the host star, and discuss scenarios that could explain both this and the planet's high density, including secular interactions with other bodies in the system or a giant impact. We note that a significantly misaligned orbit for planet c is in some degree of tension with the dynamical stability of the system, and with the fact that we see both planets in transit, though these arguments alone do not preclude such an orbit. Further observations would be highly desirable to constrain the spin-orbit angle of planet c more precisely.
\end{abstract}



\section{Introduction} \label{sec:intro}
Transiting planets around young stars are useful laboratories for studying the formation and evolution of planetary systems. This is particularly true for systems younger than 100 Myr. At such young ages, the key processes shaping the final system are at their strongest, including interactions between the planets, the disk they form in, and the host star  \citep{1997ApJ...491..856B,2012A&A...547A.111M,2019AREPS..47...67O}. In particular, information about the early dynamical history of the system is encoded in the orientation of the orbits of young planets and the spin of the host star \citep{2015ARA&A..53..409W}. Planets might undergo scattering events, collisions (i.e., giant impacts), or secular perturbations after their formation within the protoplanetary disk, which can lift their orbital inclinations \citep{2008ApJ...686..580C,2003ApJ...589..605W}. Subsequently, the orbital plane might be altered by tidal dissipation \citep[see, e.g.,][]{2016CeMDA.126..275B}. Obliquity measurements for young planets with well-measured ages are vital to constrain the key timescales for these mechanisms. Specifically, measurements for systems that are too young to have undergone significant tidal alteration of obliquity allow us to gain insight into their initial configurations \citep{Mantovan2024}. However, these measurements are still scarce. Only seven such measurements have been published to date \citep[see][for a review]{Albrecht2022}. Some practical obstacles to performing these measurements are the rapid rotation and intense magnetic activity of the host stars, which severely hamper the detection and characterisation of young planets.

The AU\,Microscopii system illustrates these points. It is one of the most interesting young multi-planet systems known to exist, and is seemingly favourable for observations because it is only 9.7\,pc away \citep{Gaia2018} and is one of the brightest M dwarfs in the sky ($V=8.7$; \citealt{2006A&A...460..695T}). As a member of the Beta Pictoris moving group \citep{1999ApJ...520L.123B}, it has a well-constrained age of $24.3_{-0.3}^{+0.3}$\,Myr \citep[e.g.,][]{2014ApJ...792...37M,2014MNRAS.445.2169M,2016A&A...596A..29M,2020A&A...642A.179M,2022A&A...664A..70G} and has long been known to host an edge-on debris disk \citep{2004Sci...303.1990K}, in which fast-moving structures of unclear origin have been imaged with SPHERE on the Very Large Telescope and \textit{Hubble} Space Telescope (HST) \citep{2015Natur.526..230B,Boccaletti2018}. 
However, the star's brightness and radial velocity (RV) have been intensively monitored and found to be highly variable and complex.
The light curve displays quasi-periodic brightness fluctuations that indicate a 4.85-day rotation period \citep{1986A&A...165..135R, hebb07}, along with frequent flares \citep{2001ApJ...554..368R}. The spin of the star appears to be closely aligned with the debris disk, based on the comparison between the inclinations of the stellar rotation axis, $i_{\rm{star}}=87.3^{+1.9}_{-2.8}$ degrees, and the disk axis, $i_{\rm{disk}}=89.4^{+0.1}_{-0.1}$ degrees \citep{2023ApJ...954...10H}.

Two close-in transiting Neptune-sized planets were discovered around AU\,Mic using photometry from the Transiting Exoplanet Survey Satellite (TESS; \citealt{Ricker+2015}).
The planets were reported by \cite{Plavchan1} and \cite{martioli1} with orbital periods of $P_b=8.46$\,d and $P_c=18.86$\,d, respectively. Ongoing monitoring of the transits using the TESS, Spitzer \citep{Werner2004} and CHaracterising ExOPlanet Satellite \citep[CHEOPS;][]{Benz1} space telescopes, and ground-based observatories, revealed significant transit timing variations (TTVs) in the system \citep{S1,S2,wittrock2022,wittrock2023}. To explain the observed TTVs, \citet{wittrock2023} invoked an additional, non-transiting planet "d" located between planets b and c, with a mass comparable to the Earth and an orbital period of $P_d=12.73$\,d. Several studies have reported masses for the two transiting planets based on optical and near-infrared RV monitoring \citep{Cale21,zicher22,donati23}. The star's rapid rotation and strong magnetic activity induce RV variations that can reach up to 1 km/s in amplitude, dwarfing the planet-induced signals and making the resulting mass measurements highly dependent on the modeling of stellar activity. In this work we adopt the mass and radius estimates from \citet{donati23} of $R_b=3.55\pm0.13\,R_\oplus$, $M_b=10.2^{+3.9}_{-2.7}\,M_{\oplus}$, $R_c=2.56\pm0.12\,R_\oplus$ and $M_c=14.2^{+4.8}_{-3.5}\,M_\oplus$. \citet{donati23} did not confirm or rule out planet d, but reported an additional candidate non-transiting planet with a period of $P_e=33.4$\,d and a mass of $\sim 35\,M_\oplus$, which may also play a role in explaining the observed TTVs.

The fact that planet c is denser than planet b ($\rho_b=1.26_{-0.43}^{+0.68} \mathrm{~g} \mathrm{~cm}^{-3}$, $\rho_c=4.7_{-1.6}^{+2.5} \mathrm{~g} \mathrm{~cm}^{-3}$; \citealt{donati23}) suggests that planet c has a more massive solid core and a smaller H/He gaseous envelope \citep{zicher22}.
This can be in tension with the general theoretical expectation that a more massive solid core should be able to accrete a larger mass of gas from the protoplanetary disk \citep{2013ApJ...775..105O}. Possible explanations include: (a) planet c could be the result of a giant impact between two outer planets in the system, or (b) planets b and c could have formed in starkly different conditions within the same disk \citep{zicher22}. 
Measuring the orientation of AU\,Mic\,c’s orbit might be helpful to differentiate between these scenarios.

When a transiting planet crosses over the disk of the host star, a portion of the blue or red-shifted starlight (due to stellar rotation) is blocked from the line of sight. This causes distortions of the disk-integrated stellar spectrum and an ``anomalous'' apparent radial velocity, which reflects the trajectory of the planet. This is the so-called Rossiter-McLaughlin (RM) effect, which allows for the measurement of the projected angle between the planet’s orbit and the stellar rotation axis, also known as the projected spin-orbit angle \citep{2000A&A...359L..13Q,2005ApJ...631.1215W}. Multiple studies \citep{Palle2020, martioli1,hirano20,addison21}, have used RM observations to constrain the projected spin-orbit angle of AU\,Mic\,b to be consistent with zero ($\lambda_b=-2.96_{-10.30}^{+10.44}$; \citealt{Palle2020}). In this work we aim to make the first same type of measurement for planet c. A misaligned orbit for planet c would suggest that the planet has undergone significant dynamical interactions after the dispersal of the gas disk, thus favouring the giant-impact scenario (a), or other type of dynamical disruption.
In contrast, an aligned orbit would indicate a more sedate dynamical history, compatible with scenario (b).

The measurement of the projected spin-orbit angle $\lambda$ from RM observations is known to be affected by significant degeneracies with other parameters that are also related to the local radial velocity of the star along the transit chord. These include the time of the midpoint of the transit, the planetary orbital inclination $i$ (or impact parameter, $b=a \cos i /R_\star$) and the star's projected rotation velocity $v \sin i_\star$ \citep[e.g.,][]{2007ApJ...655..550G}. Thanks to the wealth of existing space-based photometry and high-resolution spectroscopy, most of these parameters are relatively well-known for AU\,Mic and its transiting planets (e.g., $P_{\rm rot}=4.856\pm0.003$\,d and $v \sin i_\star= 8.5\pm0.2$\,km/s; \citealt{donati23}). However, both planets display TTVs, with TTV amplitudes of $20$--$30$ minutes reported for planet b, and $5$--$10$ minutes for planet c \citep{S1,S2,wittrock2022,wittrock2023}. These TTVs make it challenging to predict transit midpoints precisely enough to plan and interpret our new spectroscopic observations; for planet c, the most recent published observations were conducted two years before our ESPRESSO observations. We therefore arranged for simultaneous photometric observations with CHEOPS and the Next-Generation Transit Survey \citep[NGTS;][]{wheatley2018ngts} to constrain the times of the transit midpoints.

In this work, we present the first measurement of the projected spin-orbit angle for the young sub-Neptune AU\,Mic\,c, using simultaneous observations with VLT/ESPRESSO, CHEOPS, and NGTS.
The remainder of this paper is structured as follows. We present the observations and basic data reduction steps in Section \ref{sec:obsred}, and describe the steps we took to mitigate the impact of flares in Section \ref{sec:flares}. We then present our joint model of the CHEOPS photometry and ESPRESSO RVs, and the resulting constraints on the obliquity of planet c, in Section \ref{sec:joint_analysis}. We discuss the limitations and potential implications of our results, and outline our key conclusions and future plans, in Section \ref{sec:discuss}.


\section{Observations and data reduction} \label{sec:obsred}

In this section, we present the observations used in this work and the basic data reduction steps.

\subsection{CHEOPS photometry} \label{sec:cheops_obsred}
We observed the two transits of AU\,Mic\,c that occurred on the nights of the $7^{\rm th}$ and the $26^{\rm th}$ of July 2023 (UTC) using the CHEOPS space telescope \citep{Benz1} simultaneously with the ESPRESSO observations, in order to refine the mid-transit times and characterise the photometric variability of the host star. 
CHEOPS's payload consists of a 32\,cm diameter Ritchey-Chr\'{e}tien telescope and a single-CCD camera covering the wavelength range 330--1100\,nm with a field of view of 0.32\,deg$^2$. 
Each transit was covered by a CHEOPS visit\footnote{A visit is a sequence of successive CHEOPS orbits about the Earth that are devoted to observing a single target. CHEOPS revolves around the Earth every $\sim$99 minutes in a Sun-synchronous, low-Earth orbit (700\,km altitude).} of AU\,Mic\,c, taken as part of the CHEOPS Guaranteed Time Observations (GTO) programme number CH\_PR140071, were used in this work, see Table \ref{cheopsaumiclog}.  

\begin{figure*}
\centering
\includegraphics[width=1.0\textwidth]{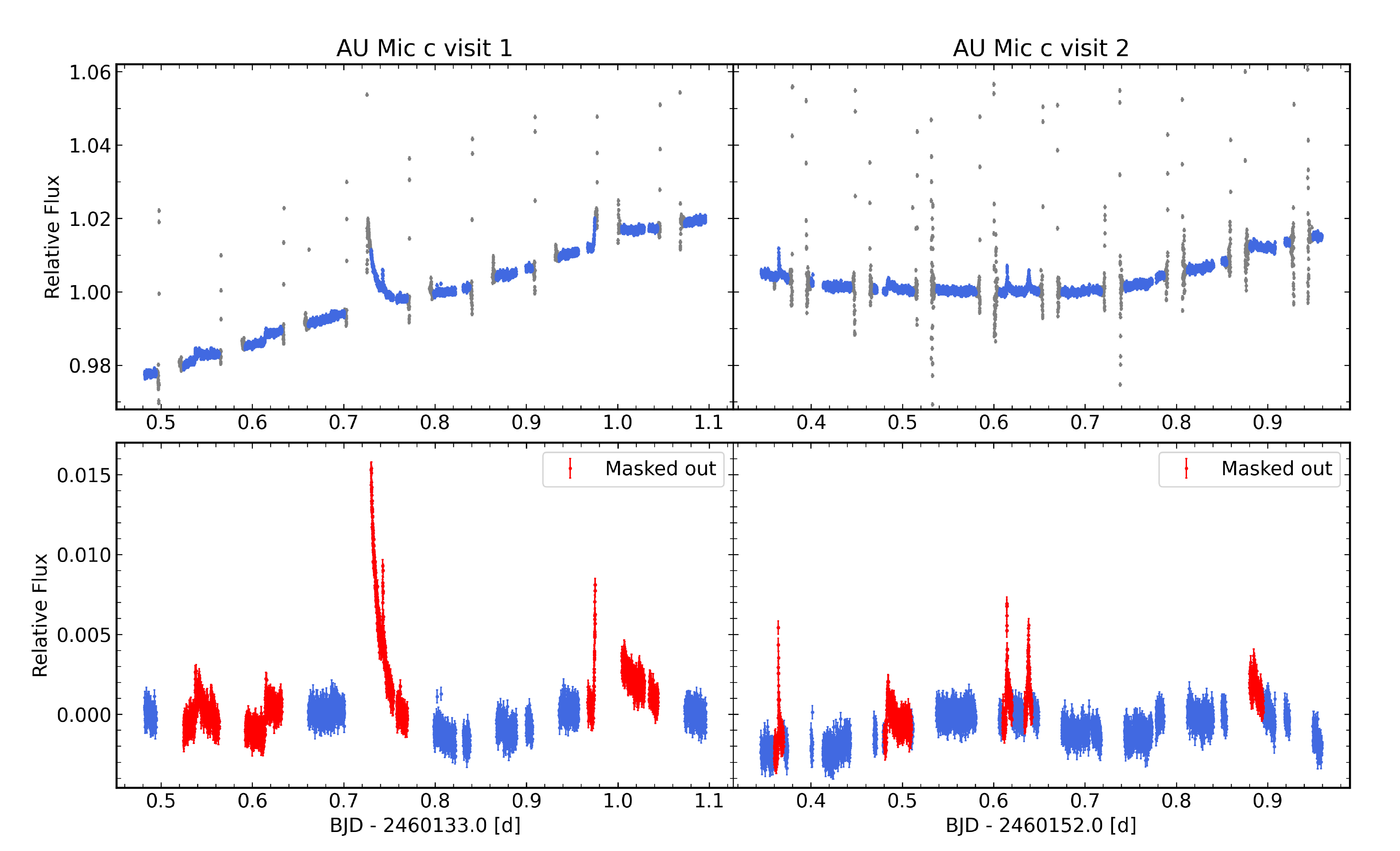}
\caption{CHEOPS photometric observations of the AU\,Mic\,c transits. The left and right column show the first and second visit, respectively. The top row shows the lightcurves as extracted by the {\tt PIPE} software after normalising each visit to unity, with the outliers caused by enhanced straylight at the start and end of each orbit marked in grey. 
The second row shows the data after subtracting a low-order polynomial trend (for visualisation purposes only), with the observations affected by flares highlighted in red. 
}
\label{fig:cheops_obs} 
\end{figure*} 

We extracted lightcurves using Point-Spread-Function (PSF) fitting on circular ``imagettes'' of 30-pixel radius centered on the target using the purpose-built software {\tt{PIPE}}\footnote{See \url{https://github.com/alphapsa/PIPE}} \citep{S1,S2,Brandeker2024}, which are downloaded for each 5-second exposure. This approach yields similar photometric precision, but better time-sampling than the standard CHEOPS {\tt{Data Reduction Pipeline}} ({\tt{DRP}})\footnote{Tested using version 14.1.3 of the DRP \citep{Hoyer1}.}. The latter performs instead aperture photometry on a $200 \times 200$-pixel subarrays of the full detector, that, due to their larger size, must be co-added in batches of 7 exposures before downloading. The top panel of Figure~\ref{fig:cheops_obs} shows the {\tt{PIPE}} CHEOPS lightcurve after normalising each visit to unity. The average uncertainty of each photometric data point in the {\tt{PIPE}} lightcurve, corresponding to a 5-second integration, is $\sim 400$\,ppm.

\begin{table*}
\centering
\caption{Log of the CHEOPS photometric observations of AU\,Mic\,c used in this work.}
\label{cheopsaumiclog}
\begin{tabular}{ccccccc}
\hline
\hline
Visit  	& Start date 			& End date 			    & File  							    & CHEOPS  		    & Integration   		& Number \\
No.    	& [UTC]      			& [UTC]    				& key   							    & product    		& time [s]      		& of frames\\   
\hline
\hline
1 		& 2023-07-07 23:25 		& 2023-07-08 14:09 		& {\tt{CH\_PR140071\_TG000901}} 		& Subarray  		& $7 \times 5.0$ 	    & 919\\
1 		& 2023-07-07 23:25 		& 2023-07-08 14:09 		& {\tt{CH\_PR140071\_TG000901}} 		& Imagettes 		& 5.0 			        & 6433\\
2 		& 2023-07-26 20:08 		& 2023-07-27 10:51 		& {\tt{CH\_PR140071\_TG000602}} 		& Subarray  		& $7 \times 5.0$ 	    & 1052\\
2 		& 2023-07-26 20:08 		& 2023-07-27 10:51 		& {\tt{CH\_PR140071\_TG000602}} 		& Imagettes 		& 5.0 			        & 7364\\
\hline
\hline
\end{tabular}
\\ \footnotesize{CHEOPS observations log. The table lists the time interval of individual observations (time notation follows the ISO-8601 convention), the file key, which can be used to locate the observations in the CHEOPS archive, the type of the photometric product, 
the integration time, and the total number of frames in each visit.}
\end{table*}

Outliers are visible just before and after each interruption within a given visit\footnote{These interruptions happen once per CHEOPS orbit when the target is occulted by the Earth, or it is barely visible due to stray light from the illuminated Earth limb or particle hits during passages above the South Atlantic anomaly.}, caused by the increased background associated with scattered light from Earth. To remove these outliers, we masked all data taken when the background flux measured by the {\tt{PIPE}} software exceeded $20\,\mathrm{electrons/pixel}$. Outliers identified in this way are marked in grey in the top panel of Figure~\ref{fig:cheops_obs}.
This resulted in the removal of 1062 out of 6433 frames (16.5\%) for the case of the first visit, and 1592 out of 7364 (21.6\%) for the second, leaving a total of 11143 data points for the remainder of the analysis.

\subsection{NGTS photometry} \label{sec:ngts_obsred}

The NGTS is a ground-based photometric facility consisting of twelve independently steerable 0.2~m diameter robotic telescopes, located at the ESO Paranal Observatory in Chile. AU\,Mic was observed by NGTS simultaneously with the ESPRESSO observations of the two transits of AU\,Mic\,c. The observations were performed using the NGTS multi-telescope mode in which multiple NGTS telescopes are used to simultaneously observe the same star. This observing mode has been shown to significantly improve the photometric precision of NGTS data \citep[e.g.][]{smith2020ngtsmulti,bryant2020multicam}. 

Nine telescopes were used on the night of the $7^{\rm th}$ of July 2023, and ten on the night of of the 26$^{\rm th}$ of July 2023. For both nights, the observations were performed using the custom NGTS filter (520--890\,nm) and an exposure time of 5\,seconds. The observations were performed with the telescopes heavily defocussed in order to avoid saturation and reduce effects of per-pixel sensitivity systematics and X-Y systematics \citep{Southworth2009a,Southworth2009b,Nascimbeni2011}. A custom aperture photometry pipeline was used to reduce the NGTS data. This pipeline uses the SEP Python library \citep{bertin96sextractor,Barbary2016} to perform the source and photometry and also utilises \textit{Gaia} DR2 information \citep{GAIA_DR2} to identify comparison stars similar to AU\,Mic in terms of brightness, colour, and CCD position. 

Both nights of observations were severely impacted by clouds, which caused large spurious fluctuations in the photometry (see top row of Figure~\ref{fig:ngts_obs}). The data from the second transit also show ramps at the beginning and end of the night, which are most likely systematic trends due to a lack of insufficient high quality comparison stars from these observations for such a bright and red star. Even after masking out the most severely affected data (middle row of Figure~\ref{fig:ngts_obs}), the transit signals cannot be easily distinguished from the strong out-of-transit trend and residual short-term systematics. Consequently, the NGTS data were not used in our final analysis, though we did compare them to our fitted model as a consistency check (see Appendix \ref{sec:ngts_analysis} for details).

\subsection{ESPRESSO spectroscopy} \label{sec:espresso_obsred}

We observed two transits of AU\,Mic\,c during the nights of July 7$^{\rm{th}}$ 2023 and July 26$^{\rm{th}}$ 2023, using the high precision spectrograph ESPRESSO \citep{Pepe2014,Pepe2021} mounted on the Very Large Telescope (VLT) (Program~ID:~111.24VF, PI:~H.~Yu). For both runs, we used one unit telescope and employed the high-resolution mode (HR; R $\sim$ 13800) and the 2x1/SLOW binning/readout mode. Each exposure lasted 300 seconds. On the first night, 85 spectra were obtained, and on the second night, 66 spectra were obtained, with mean signal-to-noise ratios (SNRs) at 550~nm of 110.2 and 115.3, respectively. During both runs, the airmass ranged between 0.7 and 1.4. During the second run, there was a problem with the control unit of the UT2 telescopes that interrupted the observations at $\mathrm{BJD}_\mathrm{TDB}=2460152.67160$. A decision was made to switch telescopes to UT1, and the observations resumed at $\mathrm{BJD}_\mathrm{TDB}=2460152.73378$. UT1 was used for the remainder of the night.

The spectra were first reduced using the official ESPRESSO data reduction software (DRS) version 3.0.0 \citep{Pepe2021}. The DRS data products include reduced 2-D and 1-D (order-merged) spectra as well as Cross-Correlation Functions (CCFs) and RVs extracted by fitting a Gaussian function to each CCFs (along with the other parameters of the Gaussian fit). A preliminary inspection of these products revealed that the spectra, CCFs and RVs were all severely affected by the flares and other stellar activity. We therefore decided to perform our own post-processing and RV extraction, as previous studies \citep[e.g.,][]{Bourrier21} have shown that this can yield significant gains over the DRS data products. 

We used the {\tt{YARARA}} package \citep{C21}, to perform post-processing of the 1D order-merged (S1D) spectra in order to remove or reduce instrumental and telluric signals. As {\tt{YARARA}} is a data-driven pipeline, it benefits from using as many observations of a given star with a given instrument as possible. We therefore also re-processed the transit of AU\,Mic\,b observed in 2019 with ESPRESSO (Program ID:~103.2010, PI:~J.~Strachan) and published by \cite{Palle2020}. This has the added benefit of broadening the range of Barycentric Earth Radial Velocity (BERV) spanned by the observations, which helps with the telluric correction step of {\tt{YARARA}}.

Before entering the {\tt{YARARA}} pipeline, the spectra were continuum subtracted using the {\tt{RASSINE}} package \citep{C20b}. Outliers due to cosmic rays or dead pixels were then masked out via sigma clipping and replaced by the time-median value. The spectra were then corrected for contamination by microtelluric lines, as described in \citet{C21}. The resulting, post-processed S1D spectra were used in the remainder of the analysis.

After combining all the spectra into a single, master-spectrum, we used this to construct a list of non-blended absorption lines \citep{C20b}. The so-called {\texttt{KITCAT}} line-list was used to compute a CCF for each spectrum, which contains in total around 5000 lines, from which RVs were extracted. However, the resulting RVs were severely affected by flares, so we later re-extracted the RVs using only the subset of lines least sensitive to the flares (see Section~\ref{sec:espresso_rvs}).

It is worth noting that the ESPRESSO and NGTS observations were taken simultaneously, and the two telescopes are only a few km apart. The clouds that affected the NGTS observations probably also affected the ESPRESSO spectra. However, we expect clouds to mainly affect the overall light level and not the shape of the spectrum and hence have a negligible impact on the extracted CCFs and RVs.

\section{Flare mitigation} \label{sec:flares}

Several flares occurred during and around each of the two observed transits of AU Mic c. Together with the longer-term trends caused by star-spots, they dominate both the CHEOPS photometry and ESPRESSO RVs. We therefore adopted the following procedure to mitigate their impact on the final analysis.

\subsection{Identifying and masking flares in the CHEOPS lightcurve} \label{sec:cheops_flares}

The impact of the flares on the CHEOPS lightcurves is easier to see after subtracting a second-order polynomial fit, intended to represent the long-term trend due to active regions rotating on the stellar surface (see bottom row of Figure \ref{fig:cheops_obs}). 

While it would in principle be possible to model the flares, this would result in a very large number of free parameters: each flare would require a minimum of three parameters (i.e., the occurrence time, amplitude, and decay rate), and there are at least 5 flares per visit. We therefore opted to mask out the data that were obviously affected by flares before modelling the spot and transit signals. The data points affected by flares were identified by visual inspection, cross-checking against the flare proxy derived from the ESPRESSO spectra (see Section~\ref{sec:espresso_flare_proxy}) whenever the two time-series overlap, and are marked in red in Figure~\ref{fig:cheops_obs}. Although the onset of a flare is relatively straightforward to identify, the tail end of a flare is much less well defined, so we opted to exclude data after each flare up to the end of the affected CHEOPS orbit. 

\subsection{Extracting a flare proxy from the ESPRESSO spectra} \label{sec:espresso_flare_proxy}

\begin{figure*}
	\centering
	\includegraphics[width=1.0\textwidth]{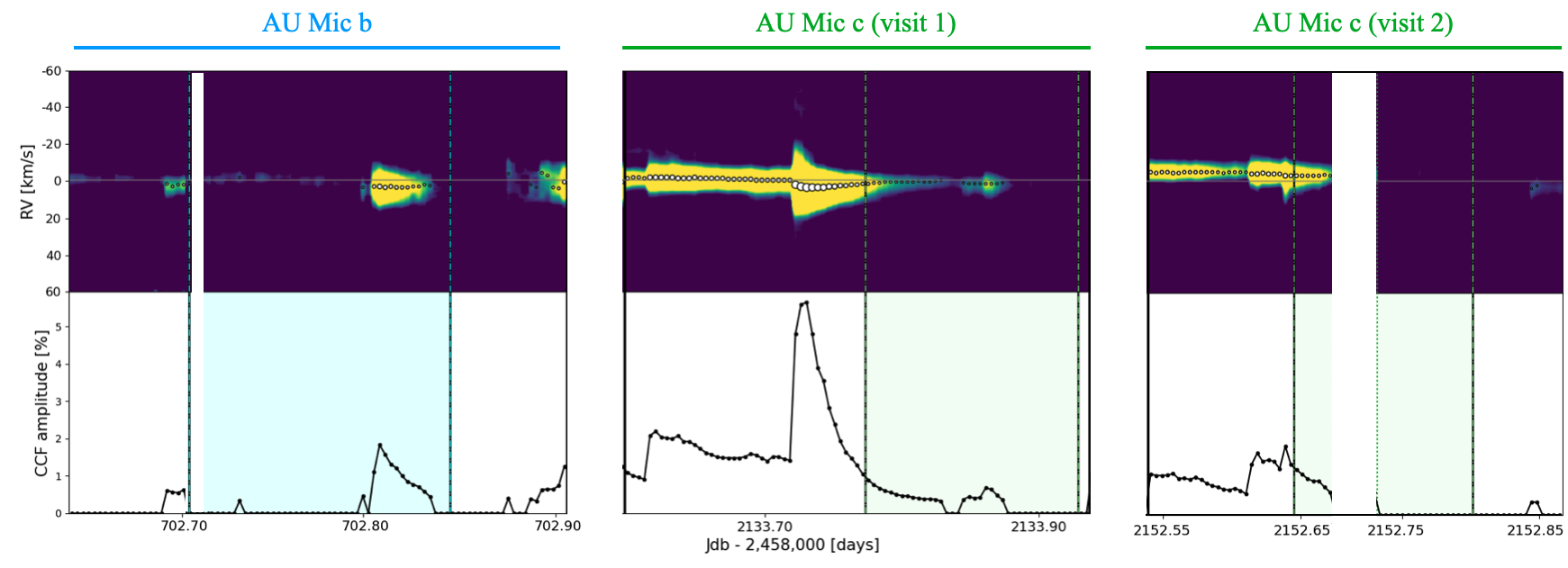}
    \caption{Extraction of the flare proxy. The top panel shows the time-series of "flare CCFs" (see text for details), from left to right: the 2021 transit of AU\,Mic\,b, then the two transits of AU\,Mic\,c observed in 2023. The white markers indicate the mean of a Gaussian fit to each CCF, while their size reflects the amplitude (or contrast of this Gaussian), which is also displayed in the bottom panel, and consists of our flare proxy. The shaded area in each panel in the bottom row indicates the transits of planets b and c. The gap during the second transit of planet c corresponds to the interruption due to the change of the UT telescope.}
    \label{fig:espresso_flare_proxy}
\end{figure*}

We first characterized the flare signal in the {\tt{YARARA}-processed} S1D spectra. The flares manifest mainly as emission lines that decay both in width and intensity after the flare impulse on a timescale of around an hour. We started by selecting two spectra occurring right before and after the start of the largest flare (occurring just before the ingress of the first transit of AU\,Mic\,c, on the night of the $7^{\rm th}$ of July 2023) and computing the difference between them. We then visually inspected this "flare spectrum" in order to identify atomic line transitions and selected a total of 33 well-defined lines between 4046 and 5434\,\AA. These lines were cross-matched with values in the National Institute for Standards and Technology (NIST) atomic spectra database\footnote{\url{https://www.nist.gov/pml/atomic-spectra-database}} to determine their rest wavelengths. The resulting line-list was then used to construct a binary mask, which was cross-correlated with each of the observed spectra, after subtracting a reference spectrum. The latter was constructed separately for each season (2021 and 2023), by averaging a small number of spectra taken well away from any flare signals. The resulting "flare CCF" time-series for the three visits is shown in the top row of Figure~\ref{fig:espresso_flare_proxy}. Each CCF in this time-series was fit with a Gaussian function, and the amplitude (or contrast) of this fit -- shown in the bottom row of Figure~\ref{fig:espresso_flare_proxy} -- constitutes our flare proxy. 

In this figure, the massive flare that occurred in the middle of the night for visit 1 of AU Mic c is clearly visible. In comparison, this flare was 5 times larger and twice as long in duration than the ones that occurred for visit 2 or the visit of AU\,Mic\,b. Despite our attempts to mitigate it, the flare occurring in this first visit was too strong and made the production of out-of-transit products impossible. It can also be noted that the RVs associated with flares were generally between $-5$ and 5 km/s, likely due to the local radial velocity of the surface where they were produced and the extra radial velocity of the ejected matter.   

\subsection{Extracting flare-resistant CCFs and RVs} \label{sec:espresso_rvs}

\begin{figure*}
	\centering
	\includegraphics[width=1.0\textwidth]{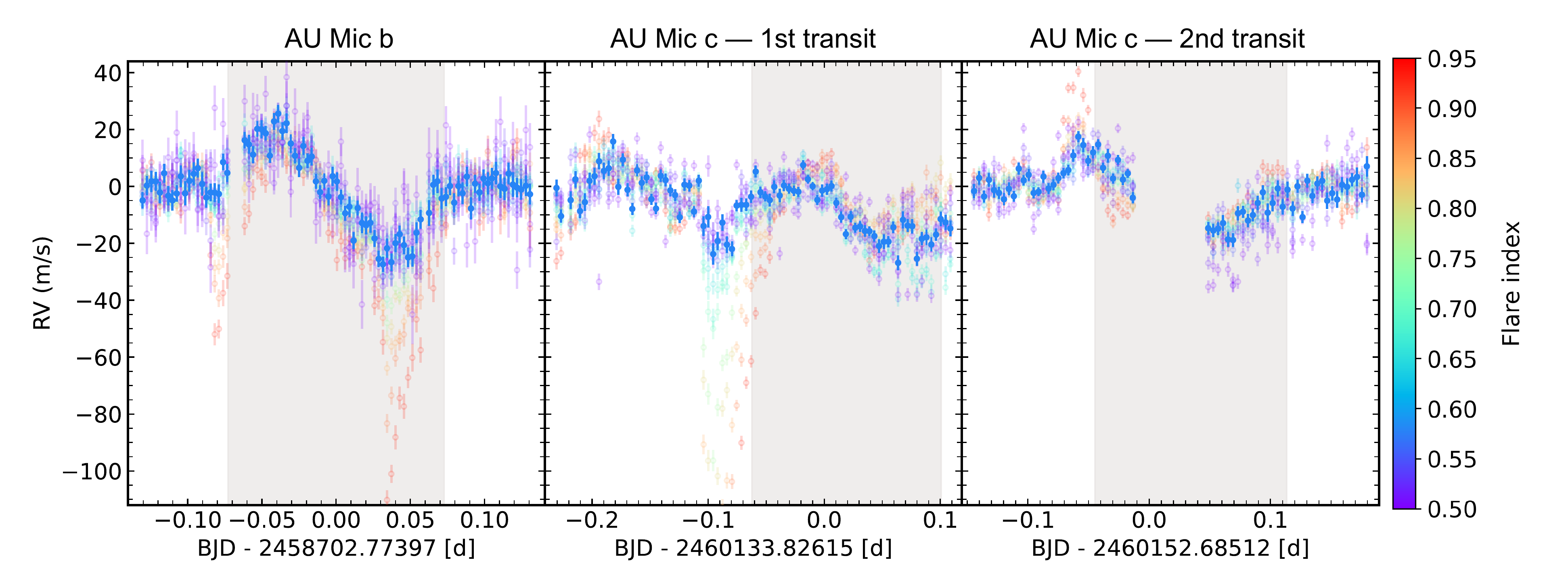}
    \caption{RV time-series extracted with progressively fewer flare sensitive lines (from red to blue) for the archival transit of planet b (left) and the two transits of AU\,Mic\,c (middle and right). The colour represents the mean flare index of all lines in the line list. Note that the flare just before the first transit of AU\,Mic\,c caused such a large RV perturbation in the RVs extracted with all the line (red points) that it goes beyond the $y$-axis range shown here. A quadratic RV trend caused by rotational modulation of active regions has been fit to the out-of-transit data and subtracted prior to plotting. The grey shaded area in each panel indicates the duration of the transit.}
    \label{fig:espresso_flare_correction}
\end{figure*}

The CCFs and RVs produced using the {\tt{KITCAT}} mask, shown in red on Figure~\ref{fig:espresso_flare_correction}, are heavily affected by the flares. We therefore sought to identify the subset of lines that are least sensitive to the flares while retaining sufficient RV information content. To quantify the flare-sensitivity of each line, we first modelled the time evolution of flux recorded in each pixel (wavelength) of the residual spectrum as the sum of a low-order polynomial (intended to account for spots on the rotating stellar surface) and a term proportional to the flare proxy derived in Section~\ref{sec:espresso_flare_proxy}. We then defined a flare index that quantifies the sensitivity of each line to flares by computing the average of the correlation coefficient over a line profile in a given velocity window of $\pm$ 10 km~s$^{-1}$ that contains most of the Doppler information of a spectral line. 

We used this coefficient to rank the lines in order of flare sensitivity, and computed CCFs and RVs with a progressively lower limiting sensitivity to flares until only the 250 least sensitive lines remained. While using fewer stellar lines decreases the impact of the flares, it increases the photon noise. By examining the transit of AU\,Mic\,b, we found that the mask with 2000 lines (out of the total $\sim$5000 lines) gave the best trade-off between flare suppression and RV precision. The RVs extracted with this mask, shown in dark blue on Figure~\ref{fig:espresso_flare_correction}, were used in the remainder of the analysis. 

The final RVs for the archival transit of planet b are very similar to those reported by \cite{Palle2020}, which strengthens our confidence in our reduction and RV extraction methodology. The RV uncertainties are somewhat larger, though, because we are using fewer lines. The flare correction described above is effective for most of the archival and new ESPRESSO observations. However, careful examination of Figure~\ref{fig:espresso_flare_correction} shows that the impact of the strongest flares on the RVs are not completely corrected when the flare proxy (displayed in the bottom row of Figure~\ref{fig:espresso_flare_proxy}) exceeds a value of 3. We therefore excluded the RVs taken at those times from the remainder of our analysis.


\section{Joint modelling of the photometry and RVs} \label{sec:joint_analysis}

We jointly modeled the CHEOPS lightcurve and the ESPRESSO flare-resistant RVs obtained during both transits of AU\,Mic\,c in order to derive constraints on the transit times and the projected orbital obliquity simultaneously, while accounting for the uncertainties arising from the star's intrinsic variability and instrumental noise. 

\subsection{The model}

For both types of observation (photometry and RVs), our model consists of three components. The first component represents the planetary transits. We used the \texttt{batman} package \citep{batman} to model the photometric transits, and the \texttt{ARoME} package \citep{2013A&A...550A..53B} to model the RM effect in the RVs (this model is optimized for CCF-derived RVs). The transit models used for both transits and both types of observations share the same star and planet parameters, including: the sky-projected stellar obliquity ($\lambda$), inclination ($i$), scaled semi-major axis ($a/R_\mathrm{s}$), planet-to-star radius ratio ($R_\mathrm{p}/R_\mathrm{s}$), and limb darkening coefficients ($u_1$ and $u_2$)\footnote{The flux-weighted mean wavelength of the customized CCF mask used for the ESPRESSO data is 624.4 nm. This closely aligns with the centre of mass of the CHEOPS passband, which ranges from 400 to 1100 nm. Therefore, we applied the same limb darkening coefficients to both the CHEOPS and ESPRESSO data.}. To account for possible TTVs, each transit was modelled using a separate mid-transit time ($T_c$). We placed uninformative uniform priors on the sky-projected stellar obliquity as well as the mid-transit times, and Gaussian priors on all other orbital parameters based on the best-fit model from the transit analysis in \citet{wittrock2023}. The RM model also depends on the projected stellar rotational velocity, $v\sin i_\star$; we placed a Gaussian prior on this parameter based on the spectroscopic measurement of \citet{donati23}.

The second component of our model consists of a second-order polynomial to account for the variations caused by active regions on the rotating stellar surface. Each dataset for each transit was modelled using a separate set of 3 polynomial coefficients. Finally, the third component is a Gaussian Process (GP) with a Mat\'ern 5/2 kernel, which absorbs short-term variations, whether caused by the star's activity or by instrumental systematics. The GP hyperparameters were shared across the two transits, but differed between the lightcurves and RVs. We adopted uninformative (broad and uniform) priors for all parameters associated with the polynomial and GP components, except for the GP timescale $\rho$ that represents how quickly the correlation between points decreases with distance, which was restricted to the range 10--60\,minutes. All the priors used are reported in Table \ref{tab:RV-param-c2-full}.
We used nested sampling with \texttt{PolyChord}\footnote{\url{https://github.com/PolyChord/PolyChordLite}}\citep{2015MNRAS.450L..61H,2015MNRAS.453.4384H} to explore the posterior distribution of the free parameters in the model. 

\subsection{Results}
\begin{figure*}
	\centering
	\includegraphics[width=1.0\textwidth]{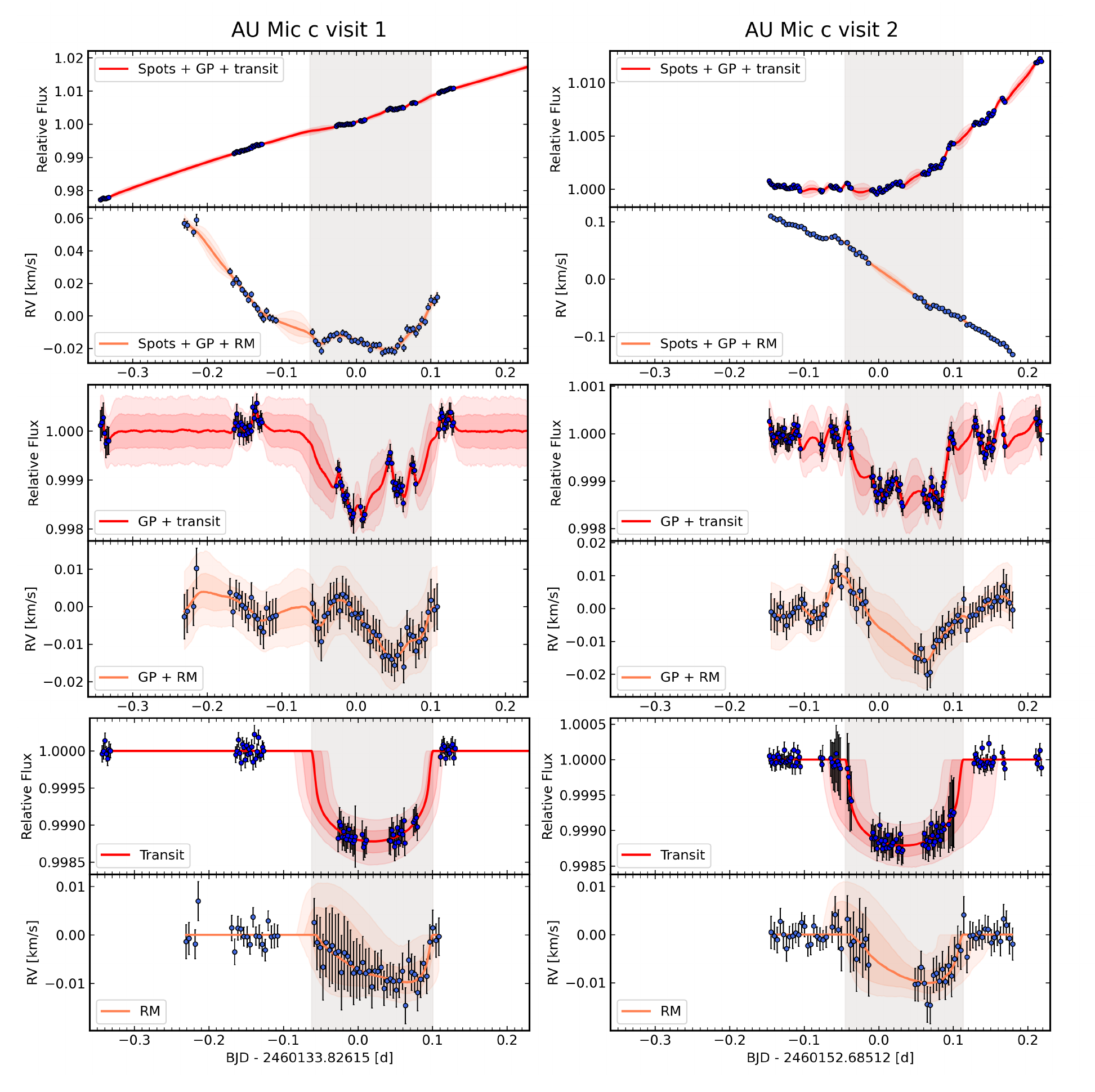}
    \caption{Joint analysis of the first and second transits of AU\,Mic\,c. The upper plots show the raw data with lines and contours corresponding to the full model (GP + polynomial + transits). The middle plots show the data and models subtracting the polynomial component. The lower plots show the model and data further subtracting the GP component. For each plot, the model is estimated through 10,000 randomly selected samples with corresponding weights generated from the nested sampling. The solid lines correspond to the weighted median of the samples, while the contours correspond to the 68.3\% (which is equivalent to 1-$\sigma$ for a Gaussian distribution) and 95$\%$ credible intervals (CIs) of the samples. Each of the samples obtained via nested sampling is associated with a weight, which was taken into account when evaluating the percentiles used to compute the credible intervals. In the middle and lower plots, the data points are the original data minus the weighted median of the samples for the components of the model subtracted, evaluated at the location of the observations, and their error bars are calculated as the quadrature sum of the original error bars and the weighted standard deviation of these samples. The solid lines and the surrounding shaded areas represent the median, 68.3\% and 95\% credible intervals of the transit model. The grey shaded area in each panel indicates the duration of the transit.}
    \label{fig:aumicc_rm}
\end{figure*}

\begin{figure*}
	\centering
	\includegraphics[width=0.9\textwidth]{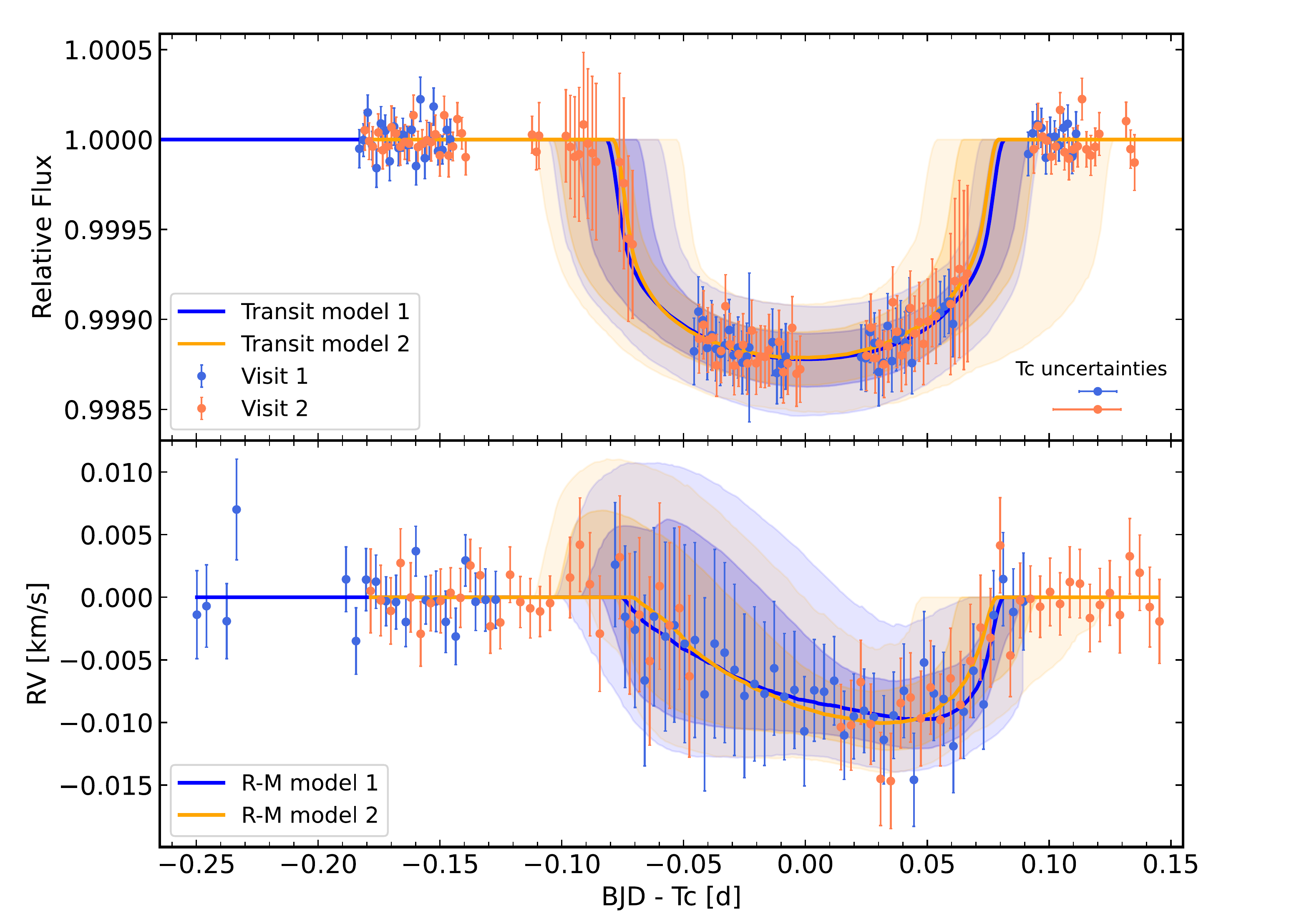}
    \caption{Phase-folded lightcurve (top) and RV (bottom) observations of the transits of AU\,Mic\,b (transit 1 in blue, transit 2 in orange). The polynomial and GP terms used to model the spot signals and residual short-term variations respectively have been subtracted from the data, and the vertical error bars incorporate the uncertainty in those model components. The solid line and shaded areas represent the median, 68.3\% ($1\sigma$) and 95\% ($\approx2\sigma$) credible intervals of the transit component of the model. The horizontal errorbars shown in the top panel represent the $1\sigma$ uncertainties on mid-transit times $Tc$ of each transit.}
    \label{fig:rv_fit_c2}
\end{figure*}

\begin{figure*}
	\centering
	\includegraphics[width=0.8\textwidth]{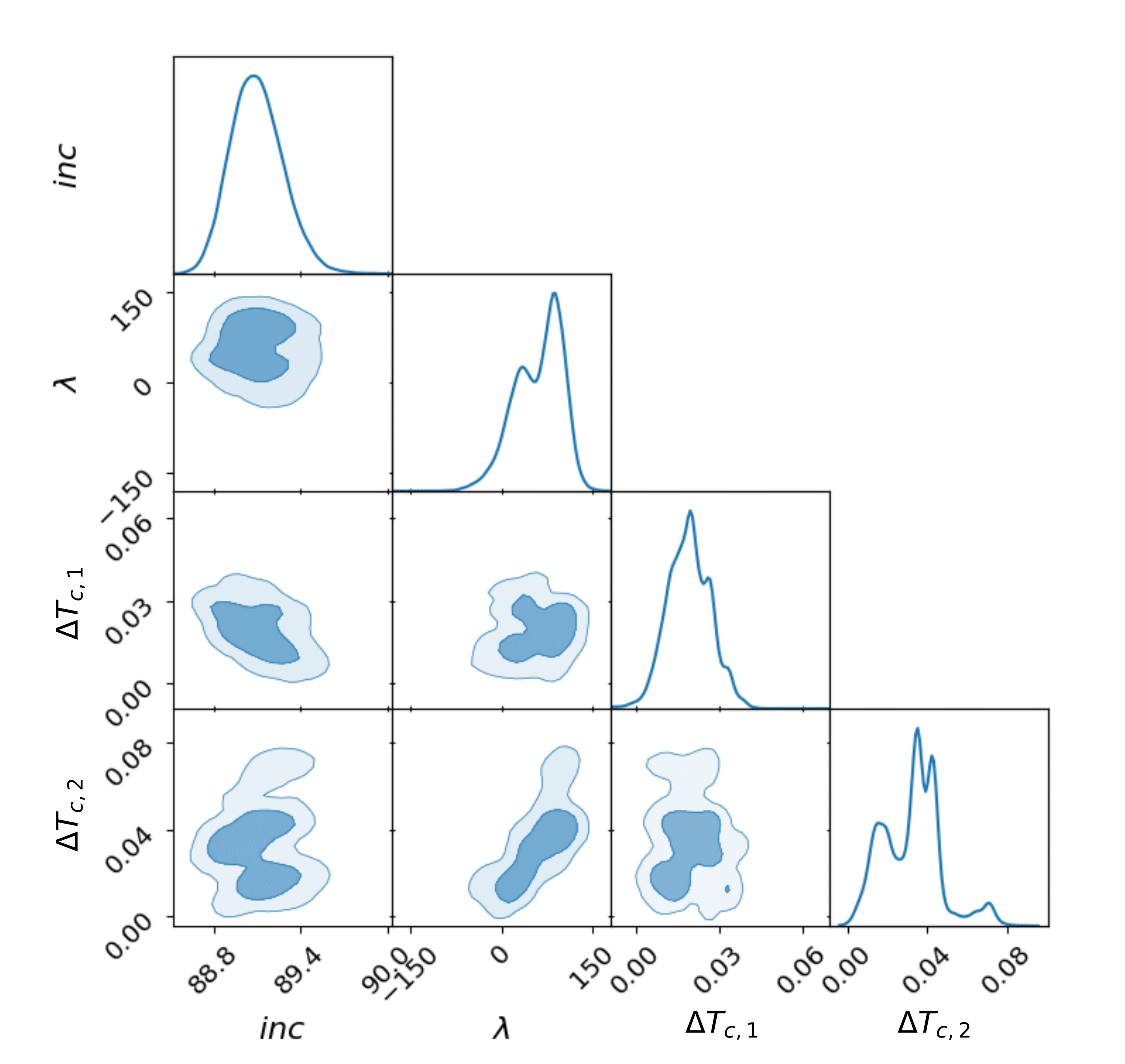}
    \caption{Posterior distribution of key parameters in the joint analysis of the first and second transits of AU\,Mic\,c.}
    \label{fig:aumicc_rm_pt}
\end{figure*}

The best-fit models are shown over-plotted on the data for both the lightcurves and RVs in Figure \ref{fig:aumicc_rm}. From top to bottom, the different rows of panels show: the input data with lines and contours corresponding to the full model (GP + polynomial + transits), the data and models after subtracting the polynomial component, and after subtracting both the polynomial and the GP components. We also show a phase-folded version of the two transits after subtracting the polynomial and GP components of the model in Figure~\ref{fig:rv_fit_c2}. The posterior distributions for the key parameters of the fit are plotted Figure~\ref{fig:aumicc_rm_pt}, while the priors and inferred 68.3\% and 95\% credible intervals in Table~\ref{tab:RV-param-c2}. The priors and inferred values of all free parameters are listed in Table \ref{tab:RV-param-c2-full}, and the full posterior distributions are shown in Figure~\ref{fig:aumicc_rm_pt-full} in the appendix. 

The best-fit model shown in the bottom-panel of Figure~\ref{fig:aumicc_rm} and in Figure~\ref{fig:rv_fit_c2} displays the one-sided RM signature characteristic of a misaligned orbit (as opposed to a prograde or retrograde one), and the best-fit (maximum a-posteriori) value of the projected spin-orbit angle is $\lambda \approx 87$\,degrees. However, the posterior distribution for $\lambda$ is bimodal, with a second subsidiary maximum at $\lambda \approx 34$\,degrees, and a width that appears to encompass 0\,degrees. Formally, we measure $\lambda=67.8_{-49.0}^{+31.7}$\,degrees, where the central value represents the median and the uncertainties encompass the central 68.3\% of the posterior distribution. After taking into account their weights from the nested sampling, 89\% of the posterior samples correspond to somewhat misaligned orbits ($\lambda_c \ge 10$\,degrees). 

The bimodality in the projected spin-orbit angle is partly explained by its strong correlation with the mid-transit times, particularly for the second transit. The latter is itself multi-modal, owing to the gaps in the CHEOPS lightcurve, and this has a knock-on effect on the spin-orbit angle. The derived mid-transit times correspond to TTVs of 27 and 49 minutes relative to the published ephemerides -- somewhat larger than the TTVs previously reported for planet c, though the uncertainties are also large ($\sim 15$ minutes at 1-$\sigma$). 

\renewcommand{\arraystretch}{1.2}
\begin{table}
    \begin{center}
	\caption{Inferred values for the key parameters from the joint analysis of the first and second transits of AU\,Mic\,c. The full set of parameters included in the fit is reported in Table \ref{tab:RV-param-c2-full}.}
	\label{tab:RV-param-c2}
	\begin{tabular}{lcrcl}
        \hline Parameter & Value$^{\mathrm{(a)}}$ & \multicolumn{3}{l}{95\% credible interval} \\ \hline
        $\lambda_c$ (deg) & $67.8_{-49.0}^{+31.7}$ & $-23.2$ -- $120.7$\\
        $i_c$ (deg) & $89.09_{-0.17}^{+0.18}$ & $88.77$ -- $89.47$\\
        $T_{c,1}-2460133.82615^{\mathrm{(b)}}$ (days) & $0.0191_{-0.0076}^{+0.0078}$ & $0.0046$ -- $0.0343$ \\
        $T_{c,2}-2460152.68512^{\mathrm{(b)}}$ (days) & $0.0341_{-0.0183}^{+0.0095}$ & $0.0073$ -- $0.0703$ \\
        \hline
        \multicolumn{5}{l}{\footnotesize $^{\mathrm{(a)}}$ Median and $68.3$\% credible interval} \\
        \multicolumn{5}{l}
{\footnotesize $^{\mathrm{(b)}}$ Transit times are quoted relative to the published ephemerides}
\\
\multicolumn{5}{l}
{\footnotesize \quad \quad  of \citep{wittrock2023}, namely $Tc=2458342.2240$ }
\\
\multicolumn{5}{l}
{\footnotesize \quad  \quad  and $P_c=18.858970$ d. }
    \end{tabular}
    \end{center}
\end{table}

\begin{figure}
	\centering
	\includegraphics[width=\columnwidth]{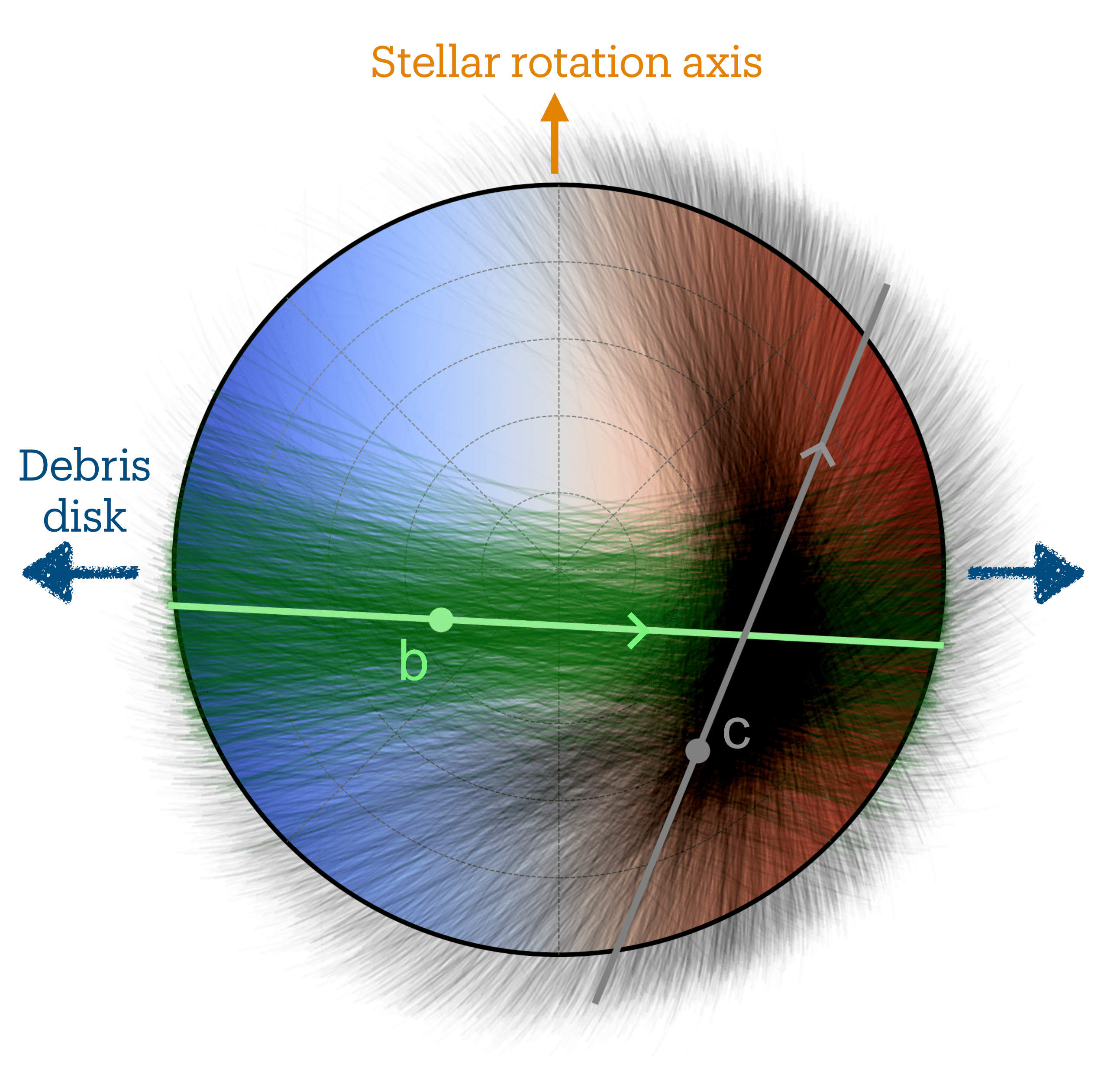}
        \caption{Schematic representation of the spin-orbit alignment of planets b and c within the AU\,Mic system. The stellar disk is colour-coded to represent local RVs, which vary as the star rotates, transitioning from blue-shifted on the left to red-shifted on the right. The bold green and grey lines represent the inferred orbits for planet b from \citet{Palle2020} and planet c from this work, with arrows indicating the directions of motion for the planets. Thin green and grey lines represent 50,000 randomly selected samples for the orbit of planets b and c, respectively. For the orbit of planet c, the transparency of the thin grey lines scales as the weight of each sample in the nested sampling. The orange arrow marks the axis of stellar rotation. The dark blue arrow marks the orientation of the debris disk.}
    \label{fig:arc}
\end{figure}

\section{Discussion and conclusions}\label{sec:discuss}

We have used simultaneous high-precision photometry from CHEOPS and precise RVs from ESPRESSO to measure the projected obliquity of AU Mic relative to the orbital plane of the young, dense sub-Neptune AU\,Mic\,c. While inconclusive, our results provide a tentative indication (89\% confidence) that the orbit of AU\,Mic\,c is misaligned with the spin axis of its host star, and hence also with planet b's orbit and the debris-disk seen at wider separation. 
Figure~\ref{fig:arc} shows a schematic representation of the orbital configuration of planet c relative to the host star and the debris disk, based on the posterior samples from our joint fit of the CHEOPS lightcurve and ESPRESSO RVs.

\begin{figure}
	\centering
	\includegraphics[width=\columnwidth]{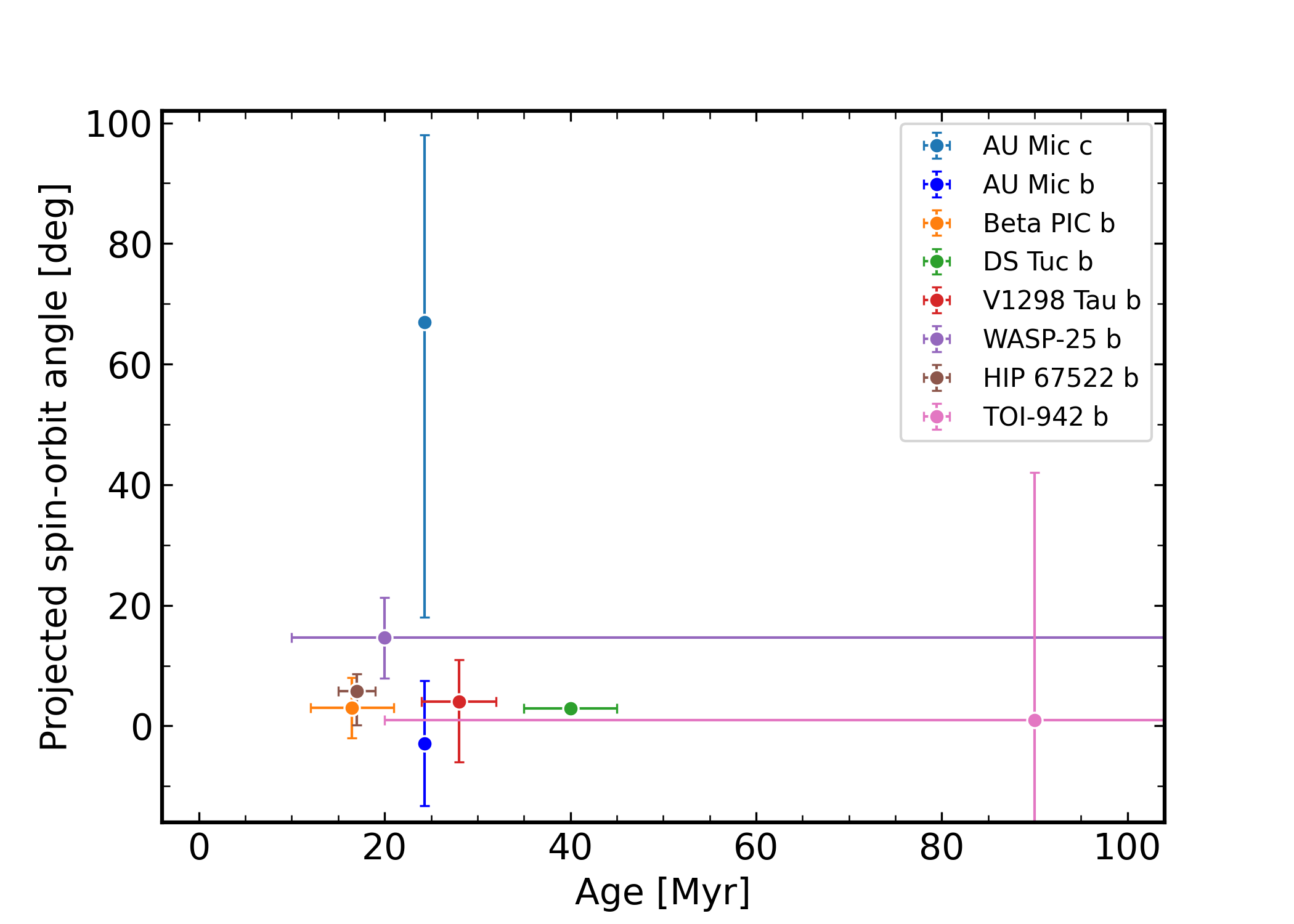}
        \caption{Projected spin-orbit angle of planets in systems younger than 100 Myr. The measurement of AU Mic c obtained in the work is shown in light blue. The measurements of other systems are from \citet{Albrecht2022}.}
    \label{fig:young_obl}
\end{figure}

If this result is confirmed, it will be the first observation of a young ($\leq$ 100 Myr) planet on a misaligned orbit. We note that, as the star is seen essentially equator-on, the true obliquity for both planets' orbits is within a few degrees of the sky-projected value. Only 7 other planets with ages $\leq$ 100 Myr have measured obliquities to date. In Figure \ref{fig:young_obl}, we show the projected spin-orbit angle versus the age of these planets, together with the measurement of AU Mic c obtained in this work. 

Moreover, very few systems with two transiting planets exhibit large mutual inclinations, irrespective of age. One comparable example is the HD\,3167 system, where the close-in super-Earth appears to be on an aligned orbit \citep{Bourrier2021}, while the outer mini-Neptune is on a nearly perpendicular orbit \citep{Dalal+2019}. 

\subsection{Limitations of our measurement}

The large uncertainties on our reported value of the projected spin-orbit angle $\lambda$ for planet c preclude any claim of a significant misalignment at this stage (which would require at least a $3\sigma$ departure from $\lambda=0$). 

Degeneracies with other parameters, such as orbital inclination $i$ and especially mid-transit times, are known to affect the measurement of the projected spin-orbit angle, $\lambda$  \citep[e.g.,][]{2010PASJ...62L..61N,2010A&A...524A..25T}. This highlights the critical role played by CHEOPS in constraining the transit times. On the other hand, even with simultaneous photometry and RVs, we were not able to constrain the transit times as well as we might have hoped given the precision and cadence of our observations, and this had a knock-on effect on the obliquity measurements. This was partly because strong flares coincided with the ingress and egress of both transits, making key parts of the data unusable (in addition to the gaps in each CHEOPS orbit caused by the occultations of the target by the Earth).

\subsection{The importance of joint modelling}
\label{sec:sequential}

Before we developed the joint model described in Section~\ref{sec:joint_analysis}, our initial attempts at modelling the data involved fitting the CHEOPS lightcurve first, then using the posteriors from that fit as priors on the RV fit. When fitting the CHEOPS lightcurves alone, we were able to model the flares explicitly, rather than masking them out, resulting in somewhat tighter constraints on the mid-transit times. The subsequent fit to the ESPRESSO RVs was restricted to the second transit observation, as the lack of post-transit baseline precluded the use of the first. This fit to the second transit yielded a nominally much more significant detection of a misaligned orbit, with $\lambda = 108.7\pm12.5$\,degrees. However, the fit of the CHEOPS lightcurves including flares implied anomalously large TTVs of order 50\,minutes -- much larger than the 5--10\,minute TTVs reported for planet c to date \citep{S1,S2,wittrock2022,wittrock2023}. In addition, there was then a significant tension (around 2-$\sigma$) between the mid-transit time obtained from the second CHEOPS lightcurve fit and from the RV fit (despite the use of the CHEOPS-based prior): the sequential approach does not guarantee a self-consistent fit between the lightcurve and the RVs. This motivated our decision to model both types of observations simultaneously, rather than sequentially, as reported in Section~\ref{sec:joint_analysis}. Doing this required masking the flares, which had the effect of considerably increasing the uncertainty on the mid-transit times, and hence on the projected spin-orbit angle, but allowed us to include the ESPRESSO observations of the first transit into the fit. The result is less precise, but in our view more robust, than the sequential approach. Nonetheless, we note here that the sequential fit supports a significantly misaligned orbit. 

Modeling the two observed transits jointly was also important, as the first ESPRESSO observation covered the full transit, but no post-transit baseline, while the gap in the second observation fell during the transit.

\subsection{Possible explanations for a misaligned orbit}

Despite the tentative nature of our result, it is interesting to speculate as to the mechanisms that could lead to a misaligned orbit for AU\,Mic\,c, without disrupting the mutual alignment between the star, disk, and planet b. 

One such mechanism is planet-planet scattering. Interactions between planets can lead to significant orbital changes. When these interactions drive relative velocities up to the escape velocity of the most massive body in the system, collisions or giant impacts become likely \citep{Goldreich2004,Schlichting2014}. Thus, these collisions typically occur at escape velocity and can result in inclination changes proportional to the ratio of the escape velocity to the local Keplerian velocity. Thus, an order of magnitude estimate of the inclination change is $\sim v_{\rm col}/v_{\rm Kep}$. Since $ v_{\rm esc}/v_{\rm Kep}\sim {1/2}$ then large inclination changes, similar to the one we have observed are plausible \citep[e.g.,][]{1996Sci...274..954R,1996Natur.384..619W,1997ApJ...477..781L,2014prpl.conf..787D}. In the event of a collision, such a scenario would likely result in planet c losing its atmosphere aligning with the abnormal density trend observed between planets b and c derived from mass measurements. However, a collision causing a large inclination is also likely to result in high eccentricity unless it occurred along a specific trajectory that resulted in high inclination and low eccentricity. Nonetheless, the eccentricity of AU\,Mic\,c is weakly constrained and may still be relatively significant. Additionally, debris from such a collision could potentially form a disk, allowing the planet to circularise. 

Another possibility is a nodal secular resonance, which can drive large misalignments \citep{Ward1976,Lai2014,Owen2017}. Given the possible presence of an exterior planetary companion \citep[e.g.,][]{wittrock2023,donati23}, this mechanism is a viable explanation for the planet c's misalignment. In this scenario, the nodal precession frequency of the exterior companion is driven by a dispersing protoplanetary disk, which slowly evolves as the disc disperses. If, during this evolution, the frequency aligns with the nodal precession frequency of planet c, the exterior planet could drive planet c into a highly inclined orbit.\citep{Petrovich2020}. In particular, this mechanism drives planets towards polar orbits, broadly consistent with the obliquity we measure. Such secular resonances do not necessarily result in large eccentricities, although they can be induced; however, they do require the system to be configured to allow the precession frequencies to become commensurate during disc dispersal.

A third possibility is Kozai-type interactions with the disk. Kozai-type dynamics cause inclination and eccentricity to oscillate in phases that are 90 degrees out of sync. Since RV monitoring of the system is consistent with a circular orbit for planet c \citep{zicher22,donati23}, we might currently be observing AU\,Mic\,c in a high inclination, low eccentricity phase. Moreover, if the event that caused the tilt in planet c’s orbit happened before the dispersal of the disk, the eccentricity might have been dampened as the planet repeatedly passed through the disk \citep[e.g.,][]{Teyssandier2013,Teyssandier2014}.

\subsection{Dynamic stability and transit probability}
The predicted sky-projected spin-orbit angle of planet c (Table~\ref{tab:RV-param-c2}) may correspond to a high mutual inclination between the two planets, which raises concerns about the dynamic stability and transit probability of the system.

\citet{Petrovich2020} suggests that large mutual inclinations could lead to eccentric instability, resulting in increased eccentricity, orbit crossings, and overall instability. Conversely, strong planet-planet interactions between planets b and c might stabilize the system, as indicated by \citet{Denham2019}, particularly in the presence of general relativity (GR) precession. Due to the significant mutual inclination, planet c would induce a nodal precession on planet b, causing variations in planet b's impact parameter and leading to transit duration variations (TDVs) as well as the TTVs that have been observed. More accurate predictions of the TTVs and TDVs, requiring detailed N-body calculations, could provide a further test of the mutual inclination of the system.

To check if the system can be stable, we performed a dynamical analysis in a similar way as for other planetary systems \citep[eg.][]{Correia_etal_2005, Correia_etal_2010}.
The system is integrated on a regular 2D grid of initial conditions around the best-fit solution from Table~5 in \citet{wittrock2023}.
Each initial condition is integrated for $5\,000$~yr, using the symplectic integrator SABAC4 \citep{Laskar_Robutel_2001}, with a step size of $5 \times 10^{-4} $~yr and GR corrections.
We then performed a frequency analysis \citep{Laskar_1990, Laskar_1993PD} of the mean longitude of planet c over two consecutive time intervals of 2\,500~yr, and determined the main frequency, $n$ and $n'$, respectively.
The stability is measured by $\Delta = \log |1-n'/n|$, the stability index, which estimates the chaotic diffusion of the orbits. A higher absolute value of the stability index indicates that the system is more stable. We note that the stability analysis here does not consider the impact of the debris disk.

\begin{figure*}
	\centering
	\includegraphics[width=0.8\textwidth]{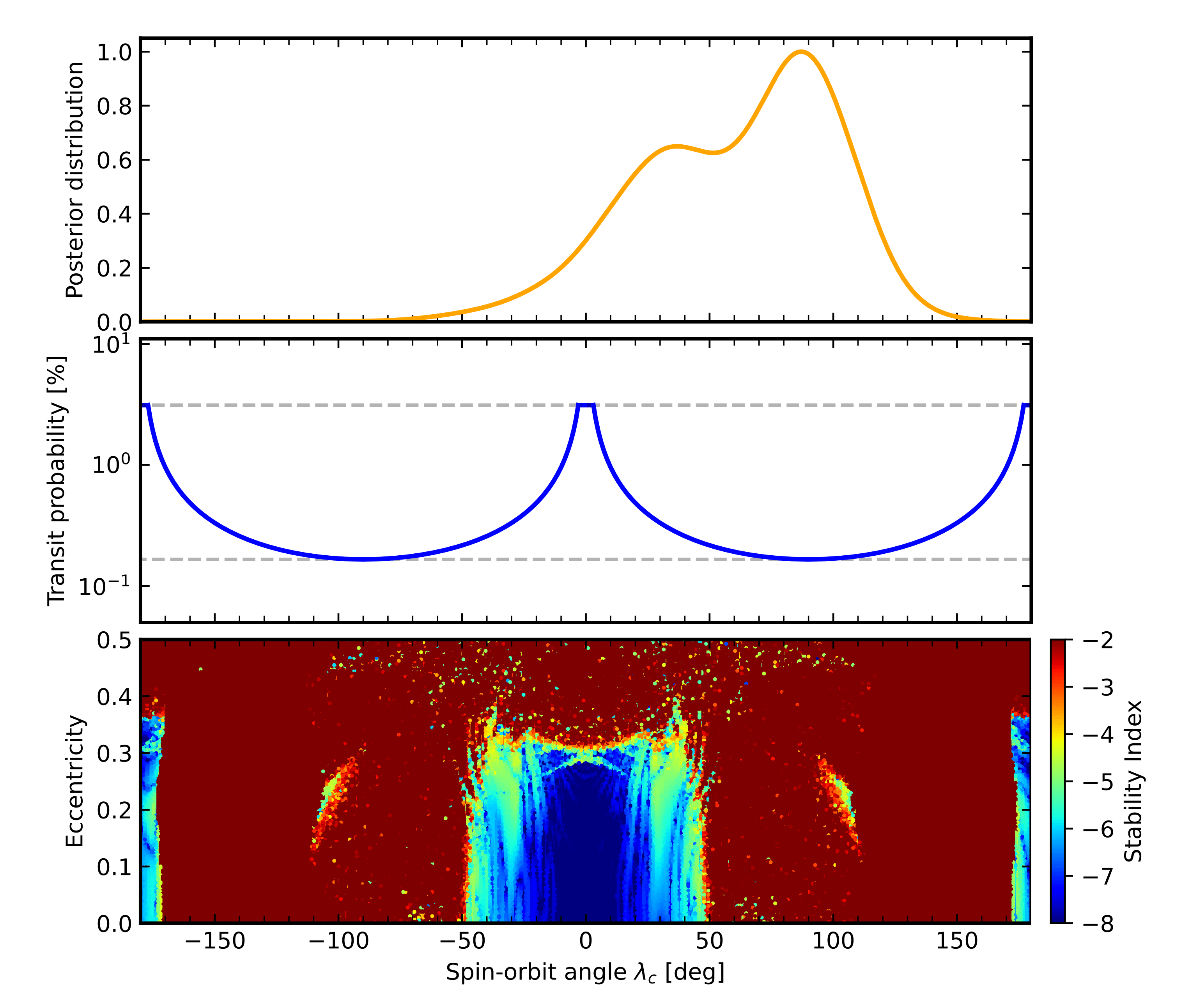}
    \caption{Upper panel: Posterior distribution of the sky-projected spin-orbit angle of AU Mic c $\lambda_c$. Middle panel: Transit probability versus $\lambda_c$, assuming AU Mic b is well aligned, estimated following \citet{Read2017}. Lower panel: Stability analysis of the AU~Mic planetary system. For fixed initial conditions (Table~5 in \citet{wittrock2023}), the parameter space of the system is explored by varying the projected spin-orbit angle and the eccentricity of planet c. The step size is $0.9^\circ$ in the spin-orbit angle and $0.0025$ in the eccentricity. For each initial condition, the system is integrated over $5\,000$~yr and a stability indicator is calculated, which involves a frequency analysis of the mean longitude of planet c. The chaotic diffusion is measured by the variation in the main frequency (see text). Red and orange points correspond to highly unstable orbits.}
    \label{fig:aumicc_da}
\end{figure*}

In the lower panel of Figure~\ref{fig:aumicc_da}, we present the stability of the system with varying the projected spin-orbit angle\footnote{We assume $i_* \approx i_b \approx i_c \approx 90^\circ$ \citep{wittrock2023}, and thus $\lambda_c \approx \Omega$ \citep{Fabrycky_Winn_2009}, where $\lambda_c$ is the spin-orbit angle and $\Omega$ is the longitude of the node of planet~c.}
and the eccentricity of planet~c. The colour represents the stability index, with orange and red indicating strongly chaotic unstable trajectories. A stability index below -4 indicates that the system is stable on time scales around 20~Myr, corresponding to the current age of the AU Mic system. 
We observe that orbits with eccentricities higher than 0.3 or spin-orbit angles higher than 50 degrees are completely unstable.

In comparison, we show the posterior distribution of the measured sky-projected spin-orbit angle of AU Mic c $\lambda_c$ in the upper panel of Figure~\ref{fig:aumicc_da}. We adopted an eccentricity for planet c of $0.041_{-0.026}^{+0.047}$ measured from RV modelling in \citet{zicher22}. The posterior distribution of $\lambda_c$ reveals a median value around $67$\,degrees and a maximum a-posteriori value around $87$\,degrees, which lies in an unstable region. However, stability is possible within the uncertainty of $\lambda_c$ for lower values, especially around the second subsidiary maximum in the posterior distribution around $34$\,degrees.

To check the transit probability of the two planets, we used the semi-analytical transit probability model presented in \citet{Ragozzine2010}, \citet{Brakensiek2016} and \citet{Read2017}. The input parameters of the model include the radius of the star and the semi-major axis of the two planets, for which we use values reported in \citet{donati23} and \citet{wittrock2023}. By assuming planet~b is on a well-aligned orbit, we estimated the double transit probability versus the projected spin-orbit angle of planet~c, shown in the middle panel of Figure~\ref{fig:aumicc_da}. Compared to the case where both planets are well-aligned, the transit probability is about 1/17 of that scenario when planet~c is misaligned with $\lambda_c = 67.8_{-49.0}^{+31.7}$\,degrees.

In summary, a significantly misaligned orbit for planet c is in some degree of tension with dynamical stability arguments, and with the fact that we see both planets in transit. While these arguments alone do not preclude such an orbit, they tend to favour the lower end of the the range of mutual inclinations allowed by our analysis and presented in Section~\ref{sec:joint_analysis}. Ideally, repeat observations should be carried out to settle the matter.

\subsection{Future work}

Given the importance of this system and the challenges caused by stellar activity, repeating the observations at a later date would be highly desirable to confirm the result and search for evidence of orbital procession. 

The enhanced activity level of AU\,Mic during our observations compared to observations of the system in 2018--2021 was evidenced by the flares and the strong spot-induced signals seen in both lightcurves and spectra. We developed a novel method to characterise the flares in the ESPRESSO data, and mitigate their impact on the extracted RVs. This technique could prove valuable for future mass and obliquity measurements of planets around active stars. 

The dataset presented in this work contains a wealth of information about the spectroscopic signature of the flares, which we intend to investigate further and report on in a future paper. 

In principle, it should also be possible to use Doppler tomography techniques to cross-check the results of our RV-analysis, as done by \citet{Palle2020} for the ESPRESSO observations of the transit of planet b. However, this would require a modelling framework capable of handling the rotating active regions present on the stellar surface during and around the transits, which is beyond the scope of the present work. 

\section*{Acknowledgements}
The authors thank Caroline Terquem, Dan Foreman-Mackey, Jiayin Dong, Fei Dai, Yu Xu for insightful discussions.
H.Y., S.A., B.K., O.B., acknowledge funding from the European Research Council under the European Union’s Horizon 2020 research and innovation programme (grant agreement No 865624, GPRV).
M.C. acknowledges the SNSF support under grant P500PT\_211024. 
H.M.C. acknowledge funding from a UKRI Future Leader Fellowship, grant number MR/S035214/1.
The contributions at the Mullard Space Science Laboratory by E.M.B. have been supported by STFC through the consolidated grant ST/W001136/1. 
J.E.O is supported by a Royal Society University Research Fellowship, and J.E.O's contribution has received funding from the European Research Council (ERC) under the European Union’s Horizon 2020 research and innovation programme (Grant agreement No. 853022).
Based on data collected under the NGTS project at the ESO La Silla Paranal Observatory.  The NGTS facility is operated by the consortium institutes with support from the UK Science and Technology Facilities Council (STFC)  projects ST/M001962/1 and  ST/S002642/1.
CHEOPS is an ESA mission in partnership with Switzerland with important contributions to the payload and the ground segment from Austria, Belgium, France, Germany, Hungary, Italy, Portugal, Spain, Sweden, and the United Kingdom. The CHEOPS Consortium would like to gratefully acknowledge the support received by all the agencies, offices, universities, and industries involved. Their flexibility and willingness to explore new approaches were essential to the success of this mission. CHEOPS data analysed in this article will be made available in the CHEOPS mission archive (\url{https://cheops.unige.ch/archive_browser/}).
Z.G. was supported by the VEGA grant of the Slovak Academy of Sciences No. 2/0031/22 and by the Slovak Research and Development Agency - the contract No. APVV-20-0148. 
Gy.M.Sz. acknowledges the support of the Hungarian National Research, Development and Innovation Office (NKFIH) grant K-125015, a a PRODEX Experiment Agreement No. 4000137122, the Lend\"ulet LP2018-7/2021 grant of the Hungarian Academy of Science and the support of the city of Szombathely. 
D.Ga. gratefully acknowledges financial support from the CRT foundation under Grant No. 2018.2323 “Gaseousor rocky? Unveiling the nature of small worlds”. 
A.C.M.C. acknowledges support from the FCT, Portugal, through the CFisUC projects UIDB/04564/2020 and UIDP/04564/2020, with DOI identifiers 10.54499/UIDB/04564/2020 and 10.54499/UIDP/04564/2020, respectively. 
A.C.M.C., A.D., B.E., K.G., and J.K. acknowledge their role as ESA-appointed CHEOPS Science Team Members. 
A.Br. was supported by the SNSA. 
M.N.G. is the ESA CHEOPS Project Scientist and Mission Representative, and as such also responsible for the Guest Observers (GO) Programme. M.N.G. does not relay proprietary information between the GO and Guaranteed Time Observation (GTO) Programmes, and does not decide on the definition and target selection of the GTO Programme. 
T.Wi acknowledges support from the UKSA and the University of Warwick. 
Y.Al. acknowledges support from the Swiss National Science Foundation (SNSF) under grant 200020\_192038. 
D.Ba., E.Pa., and I.Ri. acknowledge financial support from the Agencia Estatal de Investigación of the Ministerio de Ciencia e Innovación MCIN/AEI/10.13039/501100011033 and the ERDF “Away of making Europe” through projects PID2019-107061GB-C61, PID2019-107061GB-C66, PID2021-125627OB-C31, PID2021-125627OB-C32, PID2023-150468NB-I00, from the Centre of Excellence “Severo Ochoa” award to the Instituto de Astrofísica de Canarias (CEX2019-000920-S), from the Centre of Excellence “María de Maeztu” award to the Institut de Ciències de l’Espai (CEX2020-001058-M), and from the Generalitat de Catalunya/CERCA programme.
S.C.C.B. acknowledges support from FCT through FCT contracts nr. IF/01312/2014/CP1215/CT0004. 
L.Bo., V.Na., I.Pa., G.Pi., R.Ra., and G.Sc. acknowledge support from CHEOPS ASI-INAF agreement n. 2019-29-HH.0. 
C.Br. and A.Si. acknowledge support from the Swiss Space Office through the ESA PRODEX program. 
A.C.C. acknowledges support from STFC consolidated grant number ST/V000861/1, and UKSA grant number ST/X002217/1. 
P.E.C. is funded by the Austrian Science Fund (FWF) Erwin Schroedinger Fellowship, program J4595-N. 
This project was supported by the CNES. 
This work was supported by FCT - Funda\c{c}\~{a}o para a Ci\^{e}ncia e a Tecnologia through national funds and by FEDER through COMPETE2020 through the research grants UIDB/04434/2020, UIDP/04434/2020, 2022.06962.PTDC. 
O.D.S.D. is supported in the form of work contract (DL 57/2016/CP1364/CT0004) funded by national funds through FCT. 
B.-O.\,D. acknowledges support from the Swiss State Secretariat for Education, Research and Innovation (SERI) under contract number MB22.00046. 
This project has received funding from the Swiss National Science Foundation for project 200021\_200726. It has also been carried out within the framework of the National Centre of Competence in Research PlanetS supported by the Swiss National Science Foundation under grant 51NF40\_205606. The authors acknowledge the financial support of the SNSF. 
M.F. and C.M.P. gratefully acknowledge the support of the Swedish National Space Agency (DNR 65/19, 174/18). 
M.G. is an F.R.S.-FNRS Senior Research Associate. 
C.He. acknowledges the European Union H2020-MSCA-ITN-2019 under GrantAgreement no. 860470 (CHAMELEON), and the HPC facilities at the Vienna Science Cluster (VSC). 
K.W.F.L. was supported by Deutsche Forschungsgemeinschaft grants RA714/14-1 within the DFG Schwerpunkt SPP 1992, Exploring the Diversity of Extrasolar Planets. 
This work was granted access to the HPC resources of MesoPSL financed by the Region Ile de France and the project Equip@Meso (reference ANR-10-EQPX-29-01) of the programme Investissements d'Avenir supervised by the Agence Nationale pour la Recherche. 
M.L. acknowledges support of the Swiss National Science Foundation under grant number PCEFP2\_194576. 
P.M. acknowledges support from STFC research grant number ST/R000638/1. 
This work was also partially supported by a grant from the Simons Foundation (PI: Queloz, grant number 327127). 
N.C.Sa. acknowledges funding by the European Union (ERC, FIERCE, 101052347). Views and opinions expressed are however those of the author(s) only and do not necessarily reflect those of the European Union or the European Research Council. Neither the European Union nor the granting authority can be held responsible for them. 
S.G.S. acknowledge support from FCT through FCT contract nr. CEECIND/00826/2018 and POPH/FSE (EC). 
The Portuguese team thanks the Portuguese Space Agency for the provision of financial support in the framework of the PRODEX Programme of the European Space Agency (ESA) under contract number 4000142255. 
V.V.G. is an F.R.S-FNRS Research Associate. 
J.V. acknowledges support from the Swiss National Science Foundation (SNSF) under grant PZ00P2\_208945. 
E.V. acknowledges support from the ``DISCOBOLO'' project funded by the Spanish Ministerio de Ciencia, Innovación y Universidades undergrant PID2021-127289NB-I00. 
N.A.W. acknowledges UKSA grant ST/R004838/1.
M.L. acknowledges the support from the UKRI grant (grant number: EP/X027562/1).

\section*{Data Availability}
The photometric and radial velocity data used in this work are publicly available at CDS via anonymous ftp.

\bibliographystyle{mnras}
\bibliography{ref} 

\appendix

\section{Affiliations}
\label{sec:affiliations}

\textsuperscript{\hypertarget{inst:1}{1}} Sub-department of Astrophysics, Department of Physics, University of Oxford, Oxford OX1 3RH, UK \\
\textsuperscript{\hypertarget{inst:2}{2}} HUN-REN--ELTE Exoplanet Research Group, Szent Imre h. u. 112, H-9700 Szombathely, Hungary \\
\textsuperscript{\hypertarget{inst:3}{3}} ELTE Gothard Astrophysical Observatory, Szent Imre h. u. 112, H-9700 Szombathely, Hungary \\
\textsuperscript{\hypertarget{inst:4}{4}} Astronomical Institute, Slovak Academy of Sciences, 05960 Tatranská Lomnica, Slovakia \\
\textsuperscript{\hypertarget{inst:5}{5}} Dipartimento di Fisica, Università degli Studi di Torino, via Pietro Giuria 1, I-10125, Torino, Italy \\
\textsuperscript{\hypertarget{inst:6}{6}} Mullard Space Science Laboratory, University College London, Holmbury St Mary, Dorking RH5 6NT, UK \\
\textsuperscript{\hypertarget{inst:7}{7}} CFisUC, Departamento de Física, Universidade de Coimbra, 3004-516 Coimbra, Portugal \\
\textsuperscript{\hypertarget{inst:8}{8}} Department of Astronomy, Stockholm University, AlbaNova University Center, 10691 Stockholm, Sweden \\
\textsuperscript{\hypertarget{inst:9}{9}} Astrophysics Group, Department of Physics, Imperial College London, Prince Consort Road, London SW7 2AZ, UK \\
\textsuperscript{\hypertarget{inst:10}{10}} Department of Earth, Planetary, and Space Sciences, University of California, Los Angeles, CA 90095, USA \\
\textsuperscript{\hypertarget{inst:11}{11}} European Space Agency (ESA), European Space Research and Technology Centre (ESTEC), Keplerlaan 1, 2201 AZ Noordwijk, The Netherlands \\
\textsuperscript{\hypertarget{inst:12}{12}} Department of Astrophysical Sciences, Princeton University, Princeton, NJ 08544, USA \\
\textsuperscript{\hypertarget{inst:13}{13}} Observatoire astronomique de l'Université de Genève, Chemin Pegasi 51, 1290 Versoix, Switzerland \\
\textsuperscript{\hypertarget{inst:14}{14}} Department of Physics, University of Warwick, Gibbet Hill Road, Coventry CV4 7AL, United Kingdom \\
\textsuperscript{\hypertarget{inst:15}{15}} Centre for Exoplanets and Habitability, University of Warwick, Coventry CV4 7AL, UK \\
\textsuperscript{\hypertarget{inst:16}{16}} Konkoly Observatory, HUN-REN Research Centre for Astronomy and Earth Sciences, Konkoly Thege út 15-17., H-1121, Budapest, Hungary \\
\textsuperscript{\hypertarget{inst:17}{17}} CSFK, MTA Centre of Excellence, Budapest, Konkoly Thege út 15-17., H-1121, Hungary \\
\textsuperscript{\hypertarget{inst:18}{18}} University Observatory, Faculty of Physics, Ludwig-Maximilians-Universität München, Scheinerstr. 1, 81679 Munich, Germany \\
\textsuperscript{\hypertarget{inst:19}{19}} European Southern Observatory (ESO), Karl-Schwarzschild-Str. 2, 85748 Garching bei München, Germany \\
\textsuperscript{\hypertarget{inst:20}{20}} Center for Space and Habitability, University of Bern, Gesellschaftsstrasse 6, 3012 Bern, Switzerland \\
\textsuperscript{\hypertarget{inst:21}{21}} Space Research and Planetary Sciences, Physics Institute, University of Bern, Gesellschaftsstrasse 6, 3012 Bern, Switzerland \\
\textsuperscript{\hypertarget{inst:22}{22}} Instituto de Astrofísica de Canarias, Vía Láctea s/n, 38200 La Laguna, Tenerife, Spain \\
\textsuperscript{\hypertarget{inst:23}{23}} Departamento de Astrofísica, Universidad de La Laguna, Astrofísico Francisco Sanchez s/n, 38206 La Laguna, Tenerife, Spain \\
\textsuperscript{\hypertarget{inst:24}{24}} Admatis, 5. Kandó Kálmán Street, 3534 Miskolc, Hungary \\
\textsuperscript{\hypertarget{inst:25}{25}} Depto. de Astrofísica, Centro de Astrobiología (CSIC-INTA), ESAC campus, 28692 Villanueva de la Cañada (Madrid), Spain \\
\textsuperscript{\hypertarget{inst:26}{26}} Instituto de Astrofisica e Ciencias do Espaco, Universidade do Porto, CAUP, Rua das Estrelas, 4150-762 Porto, Portugal \\
\textsuperscript{\hypertarget{inst:27}{27}} Departamento de Fisica e Astronomia, Faculdade de Ciencias, Universidade do Porto, Rua do Campo Alegre, 4169-007 Porto, Portugal \\
\textsuperscript{\hypertarget{inst:28}{28}} Space Research Institute, Austrian Academy of Sciences, Schmiedlstrasse 6, A-8042 Graz, Austria \\
\textsuperscript{\hypertarget{inst:29}{29}} INAF, Osservatorio Astronomico di Padova, Vicolo dell'Osservatorio 5, 35122 Padova, Italy \\
\textsuperscript{\hypertarget{inst:30}{30}} Centre for Exoplanet Science, SUPA School of Physics and Astronomy, University of St Andrews, North Haugh, St Andrews KY16 9SS, UK \\
\textsuperscript{\hypertarget{inst:31}{31}} Institute of Planetary Research, German Aerospace Center (DLR), Rutherfordstrasse 2, 12489 Berlin, Germany \\
\textsuperscript{\hypertarget{inst:32}{32}} INAF, Osservatorio Astrofisico di Torino, Via Osservatorio, 20, I-10025 Pino Torinese To, Italy \\
\textsuperscript{\hypertarget{inst:33}{33}} Centre for Mathematical Sciences, Lund University, Box 118, 221 00 Lund, Sweden \\
\textsuperscript{\hypertarget{inst:34}{34}} Aix Marseille Univ, CNRS, CNES, LAM, 38 rue Frédéric Joliot-Curie, 13388 Marseille, France \\
\textsuperscript{\hypertarget{inst:35}{35}} SRON Netherlands Institute for Space Research, Niels Bohrweg 4, 2333 CA Leiden, Netherlands \\
\textsuperscript{\hypertarget{inst:36}{36}} Centre Vie dans l’Univers, Faculté des sciences, Université de Genève, Quai Ernest-Ansermet 30, 1211 Genève 4, Switzerland \\
\textsuperscript{\hypertarget{inst:37}{37}} Leiden Observatory, University of Leiden, PO Box 9513, 2300 RA Leiden, The Netherlands \\
\textsuperscript{\hypertarget{inst:38}{38}} Department of Space, Earth and Environment, Chalmers University of Technology, Onsala Space Observatory, 439 92 Onsala, Sweden \\
\textsuperscript{\hypertarget{inst:39}{39}} National and Kapodistrian University of Athens, Department of Physics, University Campus, Zografos GR-157 84, Athens, Greece \\
\textsuperscript{\hypertarget{inst:40}{40}} Astrobiology Research Unit, Université de Liège, Allée du 6 Août 19C, B-4000 Liège, Belgium \\
\textsuperscript{\hypertarget{inst:41}{41}} Department of Astrophysics, University of Vienna, Türkenschanzstrasse 17, 1180 Vienna, Austria \\
\textsuperscript{\hypertarget{inst:42}{42}} Institute for Theoretical Physics and Computational Physics, Graz University of Technology, Petersgasse 16, 8010 Graz, Austria \\
\textsuperscript{\hypertarget{inst:43}{43}} Konkoly Observatory, Research Centre for Astronomy and Earth Sciences, 1121 Budapest, Konkoly Thege Miklós út 15-17, Hungary \\
\textsuperscript{\hypertarget{inst:44}{44}} ELTE E\"otv\"os Lor\'and University, Institute of Physics, P\'azm\'any P\'eter s\'et\'any 1/A, 1117 Budapest, Hungary \\
\textsuperscript{\hypertarget{inst:45}{45}} Lund Observatory, Division of Astrophysics, Department of Physics, Lund University, Box 118, 22100 Lund, Sweden \\
\textsuperscript{\hypertarget{inst:46}{46}} IMCCE, UMR8028 CNRS, Observatoire de Paris, PSL Univ., Sorbonne Univ., 77 av. Denfert-Rochereau, 75014 Paris, France \\
\textsuperscript{\hypertarget{inst:47}{47}} Institut d'astrophysique de Paris, UMR7095 CNRS, Université Pierre \& Marie Curie, 98bis blvd. Arago, 75014 Paris, France \\
\textsuperscript{\hypertarget{inst:48}{48}} Astrophysics Group, Lennard Jones Building, Keele University, Staffordshire, ST5 5BG, United Kingdom \\
\textsuperscript{\hypertarget{inst:49}{49}} European Space Agency, ESA - European Space Astronomy Centre, Camino Bajo del Castillo s/n, 28692 Villanueva de la Cañada, Madrid, Spain \\
\textsuperscript{\hypertarget{inst:50}{50}} INAF, Osservatorio Astrofisico di Catania, Via S. Sofia 78, 95123 Catania, Italy \\
\textsuperscript{\hypertarget{inst:51}{51}} Institute of Optical Sensor Systems, German Aerospace Center (DLR), Rutherfordstrasse 2, 12489 Berlin, Germany \\
\textsuperscript{\hypertarget{inst:52}{52}} Weltraumforschung und Planetologie, Physikalisches Institut, University of Bern, Gesellschaftsstrasse 6, 3012 Bern, Switzerland \\
\textsuperscript{\hypertarget{inst:53}{53}} Dipartimento di Fisica e Astronomia "Galileo Galilei", Università degli Studi di Padova, Vicolo dell'Osservatorio 3, 35122 Padova, Italy \\
\textsuperscript{\hypertarget{inst:54}{54}} ETH Zurich, Department of Physics, Wolfgang-Pauli-Strasse 2, CH-8093 Zurich, Switzerland \\
\textsuperscript{\hypertarget{inst:55}{55}} Cavendish Laboratory, JJ Thomson Avenue, Cambridge CB3 0HE, UK \\
\textsuperscript{\hypertarget{inst:56}{56}} Institut fuer Geologische Wissenschaften, Freie Universitaet Berlin, Maltheserstrasse 74-100,12249 Berlin, Germany \\
\textsuperscript{\hypertarget{inst:57}{57}} Institut de Ciencies de l'Espai (ICE, CSIC), Campus UAB, Can Magrans s/n, 08193 Bellaterra, Spain \\
\textsuperscript{\hypertarget{inst:58}{58}} Institut d'Estudis Espacials de Catalunya (IEEC), 08860 Castelldefels (Barcelona), Spain \\
\textsuperscript{\hypertarget{inst:59}{59}} European Space Agency (ESA), European Space Operations Centre (ESOC), Robert-Bosch-Str. 5, 64293 Darmstadt, Germany \\
\textsuperscript{\hypertarget{inst:60}{60}} Space sciences, Technologies and Astrophysics Research (STAR) Institute, Université de Liège, Allée du 6 Août 19C, 4000 Liège, Belgium \\
\textsuperscript{\hypertarget{inst:61}{61}} Institute of Astronomy, University of Cambridge, Madingley Road, Cambridge, CB3 0HA, United Kingdom \\
\textsuperscript{\hypertarget{inst:62}{62}} Astrophysics Research Centre, School of Mathematics and Physics, Queen's University Belfast, Belfast, BT7 1NN, UK

\section{Table for derived values of all free parameters and their posterior distributions}

\begin{figure*}
	\centering
	\includegraphics[width=1.0\textwidth]{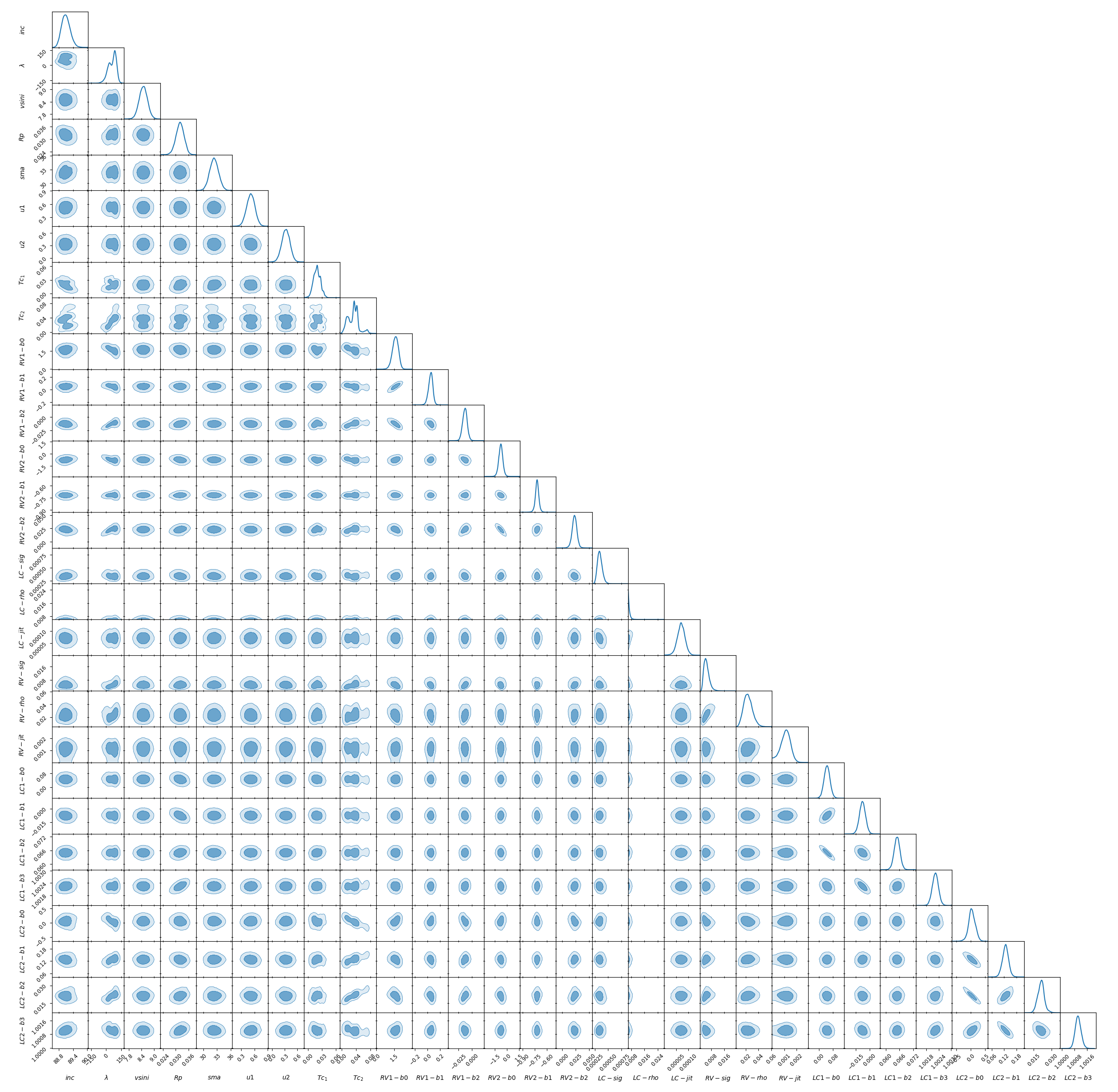}
    \caption{Posterior distribution of all free parameters in the joint analysis (CHEOPS photometry and ESRPESSO RVs) of the first and second transits of AU\,Mic\,c.}
    \label{fig:aumicc_rm_pt-full}
\end{figure*}

The parameters of the joint fit described in Section~\ref{sec:joint_analysis} are listed in Table~\ref{tab:RV-param-c2-full}, which gives the priors adopted over each parameter and the values inferred from the fit. LC-sig, LC-rho, LC-jit, RV-sig, RV-rho, RV-jit are GP hyperparameters in the model. sig is the overall variance, rho is the timescale in the Mat\'ern 5/2 kernel representing how quickly the correlation between points decreases with distance, and jit is the jitter term. The 1- and 2-D posterior distributions are shown in Figure~\ref{fig:aumicc_rm_pt-full}.

\renewcommand{\arraystretch}{1.2}
\begin{table*}
    \begin{center}
	\caption{Priors and inferred values for all free parameters from the joint analysis (CHEOPS photometry and ESRPESSO RVs) of the first and second transits of AU\,Mic\,c.}
	\label{tab:RV-param-c2-full}
	\begin{tabular}{lccrcl}
        \hline Parameter & Prior & Value$^{\mathrm{(a)}}$ & \multicolumn{3}{c}{95\% credible interval} \\ \hline
        $i$ & $\mathcal{N}[89.22,0.21]$ & $89.09_{-0.17}^{+0.18}$ & $88.77$ -- $89.47$ \\ 
        $\lambda$ & $\mathcal{U}[-180.0,180.0]$ & $67.8_{-49.0}^{+31.7}$ & $-23.2$ -- $120.7$ \\ 
        $v \rm{sin} i$ & $\mathcal{N}[8.5,0.2]$ & $8.49_{-0.19}^{+0.19}$ & $8.10$ -- $8.88$ \\ 
        $R_\mathrm{p}/R_\mathrm{s}$ & $\mathcal{N}[0.0311,0.0028]$ & $0.0320_{-0.0019}^{+0.0019}$ & $0.0281$ -- $0.0356$ \\ 
        $a/R_\mathrm{s}$ & $\mathcal{N}[32.05,1.0]$ & $32.35_{-0.95}^{+0.96}$ & $30.49$ -- $34.24$ \\ 
        $u_1$ & $\mathcal{N}[0.47,0.1]$ & $0.51_{-0.09}^{+0.09}$ & $0.33$ -- $0.70$ \\ 
        $u_2$ & $\mathcal{N}[0.30,0.1]$ & $0.32_{-0.09}^{+0.09}$ & $0.13$ -- $0.51$ \\ 
        $T_{c,1}-2460133.82615^{\mathrm{(b)}}$ (days) & $\mathcal{U}[-0.02,0.10]$ & $0.0191_{-0.0076}^{+0.0078}$ & $0.0046$ -- $0.0343$ \\ 
        $T_{c,2}-2460152.68512^{\mathrm{(b)}}$ (days) & $\mathcal{U}[-0.02,0.10]$ & $0.0341_{-0.0183}^{+0.0095}$ & $0.0073$ -- $0.0703$ \\ 
        RV1-b0 & $\mathcal{U}[-3.0,3.0]$ & $1.5982_{-0.2583}^{+0.2411}$ & $1.0643$ -- $2.0453$ \\ 
        RV1-b1 & $\mathcal{U}[-1.0,1.0]$ & $0.0540_{-0.0366}^{+0.0312}$ & $-0.0225$ -- $0.1170$ \\ 
        RV1-b2 & $\mathcal{U}[-3.0,3.0]$ & $-0.0130_{-0.0044}^{+0.0040}$ & $-0.0218$ -- $-0.0047$ \\ 
        RV2-b0 & $\mathcal{U}[-3.0,3.0]$ & $-0.7591_{-0.2538}^{+0.2436}$ & $-1.2881$ -- $-0.2426$ \\ 
        RV2-b1 & $\mathcal{U}[-1.0,1.0]$ & $-0.7207_{-0.0205}^{+0.0204}$ & $-0.7683$ -- $-0.6750$ \\ 
        RV2-b2 & $\mathcal{U}[-3.0,3.0]$ & $0.0228_{-0.0043}^{+0.0042}$ & $0.0136$ -- $0.0315$ \\ 
        LC-sig & $\mathcal{U}[0.0,0.01]$ & $0.00033_{-0.00004}^{+0.00004}$ & $0.00026$ -- $0.00044$ \\
        LC-rho & $\mathcal{U}[0.006,0.042]$ & $0.0064_{-0.0003}^{+0.0008}$ & $0.0060$ -- $0.0087$ \\ 
        LC-jit & $\mathcal{U}[0.0,0.001]$ & $0.0001_{-0.0000}^{+0.0000}$ & $0.0000$ -- $0.0001$ \\ 
        RV-sig & $\mathcal{U}[0.0,0.1]$ & $0.0050_{-0.0013}^{+0.0018}$ & $0.0028$ -- $0.0092$ \\ 
        RV-rho & $\mathcal{N}[0.018, 0.01]$ & $0.0234_{-0.0063}^{+0.0071}$ & $0.0119$ -- $0.0383$ \\ 
        RV-jit & $\mathcal{U}[0.0,0.01]$ & $0.0011_{-0.0004}^{+0.0004}$ & $0.0002$ -- $0.0018$ \\ 
        LC1-b0 & $\mathcal{U}[-2.0,2.0]$ & $0.0456_{-0.0168}^{+0.0167}$ & $0.0125$ -- $0.0795$ \\ 
        LC1-b1 & $\mathcal{U}[-2.0,2.0]$ & $-0.0078_{-0.0033}^{+0.0034}$ & $-0.0145$ -- $-0.0011$ \\ 
        LC1-b2 & $\mathcal{U}[-2.0,2.0]$ & $0.0649_{-0.0013}^{+0.0013}$ & $0.0623$ -- $0.0675$ \\ 
        LC1-b3 & $\mathcal{U}[-2.0,2.0]$ & $1.0022_{-0.0002}^{+0.0001}$ & $1.0019$ -- $1.0025$ \\ 
        LC2-b0 & $\mathcal{U}[-2.0,2.0]$ & $0.0408_{-0.1055}^{+0.1210}$ & $-0.1928$ -- $0.2729$ \\ 
        LC2-b1 & $\mathcal{U}[-2.0,2.0]$ & $0.1271_{-0.0149}^{+0.0141}$ & $0.0972$ -- $0.1560$ \\ 
        LC2-b2 & $\mathcal{U}[-2.0,2.0]$ & $0.0213_{-0.0029}^{+0.0024}$ & $0.0158$ -- $0.0271$ \\ 
        LC2-b3 & $\mathcal{U}[-2.0,2.0]$ & $1.0010_{-0.0002}^{+0.0002}$ & $1.0007$ -- $1.0014$ \\ 
        \hline
        \multicolumn{6}{l}{\footnotesize $^{\mathrm{(a)}}$ Median and $68.3$\% credible interval} \\
        \multicolumn{6}{l}
{\footnotesize $^{\mathrm{(b)}}$ Transit times are quoted relative to the published ephemerides of \citep{wittrock2023},}
\\
\multicolumn{6}{l}
{\footnotesize \quad \quad namely $Tc=2458342.2240$ and $P_c=18.858970$ d.}
    \end{tabular}
    \end{center}
\end{table*}

\section{NGTS light curve analysis} \label{sec:ngts_analysis}
We performed a stand-alone analysis of the NGTS data collected on the nights of 2023 July 7 and 26. Initially we attempted an agnostic analysis where the transit mid-point times and the planet-to-star radius ratio were allowed to vary freely. However, the NGTS data are significantly impacted by the presence of clouds during both observations. The light curves also display significant short-term variability, which may be due to flares or spots, and ramps at the beginning and the end of the observations, which are most likely due to insufficient comparison stars of a brightness and colour similar to AU~Mic. Due to this the results of this agnostic analysis are highly sensitive to the out-of-transit flux baseline model used and the priors chosen. Therefore, in this instance we feel that these NGTS observations are unable to provide independent constraints on the planet parameters or transit timing. However, the photometric precision of the NGTS measurements is high enough that, dependent on the exact transit timing, the data may provide evidence against the transit timing reported by the CHEOPS-ESPRESSO observations. We therefore perform a transit analysis to check for any inconsistency between the NGTS observations and the results of the CHEOPS-ESPRESSO joint analysis.

For this analysis, we compute the transit models using the \textsc{batman} Python package \citep{batman}. The only planet parameter we allow to vary is the transit mid-point time, \tc. For this parameter, we use a truncated Gaussian prior with a mean and standard deviation taken from the median and 68.3\% credible interval from the CHEOPS-ESPRESSO joint analysis and a strict upper limit set by the 95\% credible interval from the joint analysis. The values of these are provided in Table~\ref{tab:RV-param-c2-full}. We fix the remaining planet parameters -- orbital period, $P$, planet-to-star radius ratio, \rprs, semi-major axis scaled to the stellar radius, \ars, orbital inclination, $i$, eccentricity, $e$, argument of periastron, $\omega$, and the quadratic limb-darkening coefficients -- to the median values from the joint analysis (Table~\ref{tab:RV-param-c2-full}). Simultaneously to varying the transit mid-point time we fit the out-of-transit flux baseline using a quadratic polynomial trend in time.

\begin{figure*}
\includegraphics[width=0.95\textwidth]{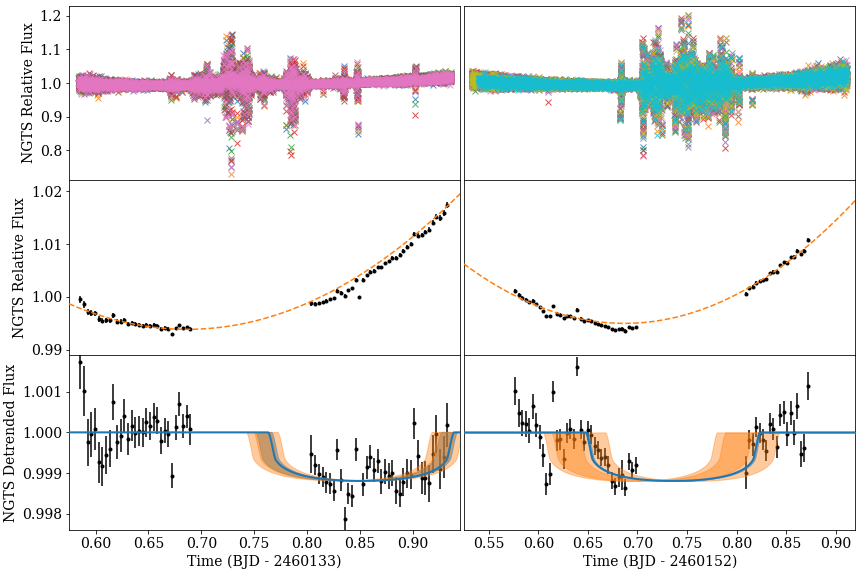}
\caption{NGTS photometric observations of the AU\,Mic\,c transits. The left and right columns show the first and second visits, respectively. The top row shows the full observations for all telescopes used, with data from each individual telescope plotted using a different colour. The middle row shows the data with the cloud-affected portion removed, and with the data from all telescopes binned together at a timescale of 5~minutes (the binning is performed for visualisation purposes only, the analysis was performed on the unbinned data). The orange dashed line in the middle row shows the best-fit polynomial component to the NGTS data. The bottom row shows the data after subtracting the polynomial component. The blue line shows the best fit model from the NGTS analysis with the blue shaded region showing the 68\% credible interval. The dark and light orange shaded regions give the 67.3\% and 95\% credible intervals respectively from the CHEOPS-ESPRESSO joint analysis.}
\label{fig:ngts_obs} 
\end{figure*}

We explored the joint posterior over the fit parameters using Markov Chain Monte Carlo (MCMC) using the \textsc{emcee} \citep{emcee} package. Twelve walkers were run for a burn-in phase of 3000 steps followed by a further 15000 steps "production" run. The resulting parameters from this analysis are provided in Table~\ref{tab:ngts} and the best-fit models are shown in Figure~\ref{fig:ngts_obs}. The resulting transit mid-point times from this analysis are \NtcNightOne\,BJD-TDB and \NtcNightTwo\,BJD-TDB for visits one and two respectively. For the night 2023 July 7 the NGTS data are fully consistent with the results of the CHEOPS-ESPRESSO joint analysis. For the night 2023 July 26 the NGTS observations suggest a slightly later transit than the 68.3\% credible interval of the CHEOPS-ESPRESSO joint analysis but fully support the presence of the transit occuring during the 95\% credible interval.

\begin{table*}
    \begin{center}
    \caption{Results obtained from the analysis of the NGTS observations. The parameter values are the median posterior values and the uncertainties give the 16 and 84 percentile ranges of the posteriors}
    \label{tab:ngts}
    \begin{tabular}{lccc}
    Name & Symbol & \multicolumn{2}{c}{Value} \\
     & & 2023 July 07 & 2023 July 26 \\
    \noalign{\smallskip} 
    \hline
    \noalign{\smallskip} 
    Transit Mid-Point Time & \tc  & \NtcNightOne & \NtcNightTwo \\
    \noalign{\smallskip} 
    Constant Detrending Coefficient & $c_{\rm 0}$ & \NcConNightOne & \NcConNightTwo \\
    \noalign{\smallskip}
    Quadratic Detrending Coefficient & $c_{\rm 1}$ & \NcLinNightOne & \NcLinNightTwo \\
    \noalign{\smallskip}
    Quadratic Detrending Coefficient & $c_{\rm 2}$ & \NcQuadNightOne & \NcQuadNightTwo \\
    \noalign{\smallskip}
    \hline    
    \end{tabular}
    \end{center}
\end{table*}

\bsp	
\label{lastpage}
\end{document}